\documentclass[reprint,aps,pra,twocolumn]{revtex4-2}

\usepackage{amsmath}
\usepackage{amsfonts}
\usepackage{amssymb}
\usepackage{amsthm}
\usepackage{mathrsfs}
\usepackage{graphicx}
\usepackage{hyperref}
\usepackage{xcolor}
\usepackage{verbatim}
\usepackage{array}
\begin{document}

\title{Three statistical descriptions of classical systems\\and their extensions to hybrid quantum-classical systems}

\author{Andr\'{e}s Dar\'{i}o Berm\'{u}dez Manjarres$^1$}
 \email{ad.bermudez168@uniandes.edu.co}
\author{Marcel Reginatto$^2$} 
\email{marcel.reginatto@ptb.de}\thanks{These authors contributed equally.}
\author{Sebastian Ulbricht$^{2,3}$}  
 \email{sebastian.ulbricht@ptb.de}
\thanks{These authors contributed equally.}
\affiliation{$^1$Universidad Distrital Francisco Jos\'{e} de Caldas, Cra. 7 No. 40B-53, Bogot\'{a}, Colombia\\
$^2$Physikalisch-Technische Bundesanstalt PTB, Bundesallee 100, 38116 Braunschweig, Germany\\
$^3$Institut für Mathematische Physik, Technische Universität Braunschweig, Mendelssohnstraße 3, 38106 Braunschweig, Germany}

\date{\today}

\begin{abstract}
We present three statistical descriptions for systems of classical particles and consider their extension to hybrid quantum-classical systems. The classical descriptions are ensembles on configuration space, ensembles on phase space, and a Hilbert space approach using van Hove operators which provides an alternative to the Koopman-von Neumann formulation. In all cases, there is a natural way to define classical observables and a corresponding Lie algebra that is isomorphic to the usual Poisson algebra in phase space. We show that in the case of classical particles the three descriptions are equivalent and indicate how they are related. We then modify and extend these descriptions to introduce hybrid models where a classical particle interacts with a quantum particle. The approach of ensembles on phase space and the Hilbert space approach, which are novel, lead to equivalent hybrid models, while they are not equivalent to the hybrid model of the approach of ensembles on configuration space. Thus, we end up identifying two inequivalent types of hybrid systems,  making different predictions, especially when it comes to entanglement.  These results are of interest regarding ``no-go'' theorems about quantum systems interacting via a classical mediator  which address the issue of whether gravity must be quantized. Such theorems  typically require assumptions that make them model dependent. The hybrid systems that we discuss provide concrete examples of inequivalent models that can be used to compute simple examples to test the assumptions of the ``no-go'' theorems and their applicability.
\end{abstract}

\maketitle

\section{Introduction}\label{secIntro}

The description of interactions between classical and quantum systems is non-trivial. First of all, it is necessary to define a common mathematical framework that is general enough to include both classical and quantum systems and allows for a sufficiently large class of interactions between them. Furthermore, one needs to choose which aspects of the classical and quantum subsystems are  considered to be essential and should be preserved in a joint interacting hybrid system, and which consistency conditions are required. Not surprisingly, different models are possible, depending on how these issues are handled, and many proposals are available in the literature (see \cite{terno2023classicalquantum} and the introduction of \cite{Barcelo_2012} for a discussion of different models and a large list of references). Up until now, none of the proposed quantum-classical models is free of difficulties and there is no consensus about the best strategy to  develop a satisfactory general theory of hybrid systems.

Despite the difficulties, there are compelling reasons to look for a general model of quantum-classical interactions. For example: (1) to explore the possibility of new physics at mesoscopic scales \cite{Barcelo_2012}, (2) to describe the measurement of a quantum system by a classical apparatus \cite{Sudarshan1976, Sudarshan1978, Sudarshan1979a, Sudarshan1979b, Reginatto_2022, Katagiri_2020}, (3) to describe the interaction between a quantum system and a classical gravitational field \cite{Boucher1988, HallReginatto2018, Oppenheim2022}, and (4) to develop better approximation techniques for complex quantum systems \cite{Burghardt2021, Burghardt2022, Melezhik2023, Gardner2023, Villaseco_Arribas2023}\footnote{We have chosen to restrict to a short list of recent references because the number of works that use quantum-classical models to approximate purely quantum systems is too vast for us to try to be comprehensive.}.

Since different hybrid models are based on different assumptions, it is of interest to compare their predictions and to investigate under what circumstances they lead to equivalent descriptions and whether there exist conditions under which their solutions can be mapped to each other. These are open questions that so far have not been systematically studied (though there are exceptions for some particular cases \cite{Zhan2008}). It is one of our goals to establish this for the three particular models that we focus on in this paper. The first model is the approach of ensembles on configuration space \cite{HallReginatto2016}. We also examine in detail the approach of ensembles on phase space, a model that was proposed recently \cite{Bermudez_Manjarres2023phase}; however, without a detailed examination of its properties and solutions. Here the main properties necessary for practical applications are studied in detail. Our third model is a Hilbert space formulation which shares many common features with the ensembles on phase space approach and is novel in that it differs from previous formulations in the assumptions made about the states and in the operator representation of the observables and generators of transformations. In particular, we show that there are general solutions where the expectation values of the operators of classical observables are equal to the corresponding classical average values (so there is no need after all to distinguish between observables and generators).  

The paper is structured as follows. In the next section, we consider three statistical descriptions of classical systems of particles and show their equivalence. We give the functional formulation of ensembles on configuration space in section \ref{ECS}, of ensembles on phase space in section \ref{EPS}, and describe the Hilbert space approach in section \ref{vHHS}. For each of them we provide the equations of motion, the definition of observables (and their Lie algebra), and discuss their relation to standard statistical mechanics formulated via the Liouville equation. In section \ref{GalInv} we consider the realization of the Galilean symmetry in each model, as this symmetry will later be required for the construction of hybrid models. We end the section by discussing the equivalence between these three classical models.

In sections \ref{HECS}, \ref{HEPS}, and \ref{HvHHS} we extend the above formalisms to quantum-classical hybrid systems, choosing the interaction term so that the hybrid system is Galilean invariant. We then examine in section \ref{comparison} under which conditions the three hybrid descriptions are equivalent. We find that the ensembles on phase space and Hilbert space approaches lead to equivalent descriptions of hybrid systems. However, they are equivalent to the hybrid formulation of ensembles on configuration space only for some special cases. 

Finally, in section~\ref{sec:discussion} we give a summary of our results and discuss some implications. 

\section{Three statistical models of interacting classical particles}

It is convenient to consider classical particles before we extend our discussion to hybrid classical-quantum systems. This also gives us the opportunity to develop much of the formalism and concepts needed later to describe hybrid models as an extension of the classical ones. Furthermore, the description of classical systems in the approaches that we present here is in itself a topic of interest. 

We consider three models of classical systems that make use of two different mathematical formalisms. The first two models are based on a probability density defined over either configuration space \cite{HallReginatto2016} or phase space \cite{Bermudez_Manjarres2023phase}. They both introduce particle dynamics via a Hamiltonian formulation of fields (where the probability density is one of two conjugate fields). The third model takes as its starting point a unitary representation of the group of contact transformation. This approach, based on the work of van Hove \cite{VanHove1951}, leads to a particular Hilbert space formulation in which the classical theory is essentially reformulated in the mathematical language of quantum mechanics. 

These three models differ from the usual descriptions of particles in classical mechanics. Instead of points in phase space (as in a Hamiltonian description) or in configuration space (as in a Lagrangian or a Hamilton-Jacobi description), the first two models utilize  probability distributions, while the third one is formulated in therms of complex probability amplitudes. Thus the models are statistical ones. However, as we will show, there is obviously a close connection between these approaches and the usual descriptions of particles in classical mechanics.

\subsection{Ensembles on configuration space}
\label{ECS}

In the approach of ensembles on configuration space \cite{HallReginatto2016}, we introduce a configuration space with coordinates $\mathbf{q}$ and define a probability density $P(\mathbf{q},t)$ which describes the uncertainty of a single classical particle's location at time $t$. The probability density must satisfy $P \geq 0$ and $\int d\mathbf{q} \,P(\mathbf{q},t) = 1$. 

\subsubsection{Equations of motion}

To derive the equations of motion, we introduce an {\it ensemble Hamiltonian} functional $\mathcal H[P,S]$, where $S=S(\mathbf{q},t)$ is an auxiliary field canonically conjugate to $P$. The physical interpretation of $S$ will be discussed below when observables are introduced. The equations of motion take the form
\begin{eqnarray}
	\frac{\partial P}{\partial t} &=& \left\{ P,{\mathcal H} \right\}_{(P,S)}  = \frac{\delta{\mathcal H}}{\delta S}\\
	\frac{\partial S}{\partial t} &=& \left\{ S,{\mathcal H} \right\}_{(P,S)}  =-\frac{\delta{\mathcal H}}{\delta P},
\end{eqnarray}
where the brackets with subscript $(P,S)$ denote the Poisson bracket of functionals with respect to $P$ and $S$; i.e., $\{ A,B \}_{(P,S)}=\int d\mathbf{q}\; \left(\frac{\delta A}{\delta P}\frac{\delta B}{\delta S} - \frac{\delta A}{\delta S}\frac{\delta B}{\delta P}\right)$ for two functionals $A[P,S]$ and $B[P,S]$. 

For the case of a single non-relativistic particle subject to a potential $V(\mathbf{q})$, the classical ensemble Hamiltonian is given by
\begin{equation}
	{\mathcal H}_{C}[P,S] = \int d\mathbf{q}\; P \left( \frac{|\nabla S|^2}{2M} + V\right), 
\end{equation}
which leads to the equations of motion
\begin{equation}
	\frac{\partial S}{\partial t} = -\frac{|\nabla S|^2}{2M}  + V, \qquad \frac{\partial P}{\partial t} = -\nabla \cdot \left( P\frac{\nabla S}{M} \right). \label{CE_HJE_ECS}
\end{equation}
The first equation is the Hamilton-Jacobi equation, the second is the continuity equation that ensures that the probability is conserved. One can show that the condition $P \geq 0$ is satisfied at all times if is initially satisfied \cite{HallReginatto2016}.

\subsubsection{Observables, generators and ensemble averages}

We associate a \textit{classical observable} ${\mathcal O}_{F}[P,S]$ to any function of phase space $F(\mathbf{q},\mathbf{p})$ by defining
\begin{equation}
	{\mathcal O}_{F}=\int d\mathbf{q}\,P(\mathbf{q})F(\mathbf{q},\nabla S).\label{classicalO}
\end{equation}

As there is an algebra of observables defined via the Poisson brackets $\{\cdot , \cdot\}_{(P,S)}$, all observables of this form also play the role of generators of continuous  transformations. Notice that the observables that correspond to average values of the derivatives of $S$ represent local energy and momentum densities, which gives a physical interpretation to the most important quantities associated with $S$ \cite{HallReginatto2016}. To see this, we calculate
\begin{equation}
	{\mathcal H}_{C} = \int d\mathbf{q} \,P\,\frac{\delta	{\mathcal H}_{C}}{\delta P} = - \int
	d\mathbf{q}\, P\,\frac{\partial S}{\partial t} = - \langle \partial S/\partial
	t \rangle ,
\end{equation}
which shows that $-P\,\partial S/\partial t$ is a local energy density. Furthermore, $\int d\mathbf{q}\, P\nabla S$ is the canonical infinitesimal generator of translations, since
\begin{eqnarray}
	\delta P(\textbf{q}) = \delta \textbf{q} \cdot \left \{ P, \int d\mathbf{q}\,
	P\nabla S \right \}_{(P,S)} = - \delta \textbf{q} \cdot \nabla P ,
	\\
	\delta S(\textbf{q}) = \delta \textbf{q} \cdot \left \{ S, \int d\mathbf{q}\,
	P\nabla S \right \}_{(P,S)} = - \delta \textbf{q} \cdot \nabla S ,
\end{eqnarray}
under the action of the generator such that $P\nabla S$ can be considered a local momentum density \cite{HallReginatto2016}. 

Finally, in the Hamilton-Jacobi theory, the momentum $\mathbf{p}$ is related to $S$ by $\mathbf{p}=\nabla S$, thus Eq. (\ref{classicalO}) implies that \textit{the numerical value of a classical observable ${\mathcal O}_{F}$ may be associated with the ensemble average of $F(\mathbf{q},\mathbf{p})$.}

\subsubsection{Observables and conservation of probability}
\label{ECSobsProb}

Changes in $P$ induced by observables ${\mathcal O}_{F}$ must preserve both the normalization and positivity of the probability. We show now that this is indeed the case \cite{HallReginatto2016}.

The infinitesimal transformation induced by an observable ${\mathcal O}_{F}[P,S]$ on $P$ is given (for infinitesimal $\epsilon$) by
\begin{equation}
	\delta P = \epsilon \{ P,{\mathcal O}_{F} \}_{(P,S)} = \epsilon \frac{\delta {\mathcal O}_{F}}{\delta S}.
\end{equation}
Thus, the normalization of the probability will be preserved if $\int d\mathbf{q}\; (P+\delta P) = 1$ or, equivalently, if $\int d\mathbf{q}\; \delta P = 0$. We have
\begin{eqnarray}
	\int d\mathbf{q}\; \delta P &=&  \int d\mathbf{q}\; \epsilon\frac{\delta {\mathcal O}_{F}}{\delta S}  \\
 &=& \int d\mathbf{q}\; \left({\mathcal O}_{F}[P,S+\epsilon]-{\mathcal O}_{F}[P,S]\right)=0\nonumber
\end{eqnarray}
as required, where the last equality follows because ${\mathcal O}_{F}[P,S]$ only depends on derivatives of $S$, as can be seen from Eq. (\ref{classicalO}).
Furthermore, 
\begin{eqnarray}
	\frac{\delta {\mathcal O}_{F}}{\delta S} &=& 0 ~~ \mathrm{if} ~~ P(\mathbf{q})=0,
\end{eqnarray}
ensures that the transformation generated by the observable ${\mathcal O}_{F}$ will preserve the positivity of $P$ \cite{HallReginatto2016}. This result follows from
\begin{eqnarray}
	\frac{\delta {\mathcal O}_{F}}{\delta S} &=& -\nabla \cdot \left[ P \, \frac {\partial F(\mathbf{q},\nabla S)} {\partial \nabla S} \right] \\
	&=& -P \, \nabla \cdot \left[\frac {\partial F(\mathbf{q},\nabla S)} {\partial \nabla S} \right] -\nabla P \cdot \left[  \frac {\partial F(\mathbf{q},\nabla S)} {\partial \nabla S} \right] = 0,\nonumber
\end{eqnarray}
where the last equality holds because $P$ is non-negative, thus it must reach a minimum for any point $\mathbf{q'}$ at which $P(\mathbf{q'})=0$, which in turn implies that $\nabla_qP|_{\mathbf{q}=\mathbf{q'}} = 0$.

\subsubsection{Lie algebra of observables}

The fundamental importance of the definition of classical observables ${\mathcal O}_{F}[P,S]$ of Eq. (\ref{classicalO}) derives from the fact that the Poisson bracket for classical observables is \textit{isomorphic} to the phase space Poisson bracket \cite{HallReginatto2016},
\begin{equation}
	\left\{ {\mathcal O}_{A},{\mathcal O}_{B}\right\}_{(P,S)}={\mathcal O}_{\left\{ A,B \right\}_{(q,p)} },\label{iso1}
\end{equation}
where on the right side we have introduced the usual Poisson bracket in phase space; i.e., $\{A,B\}_{(q,p)}=\sum_k \left(\frac{\partial A}{\partial q_k}\frac{\partial B}{\partial p_k} - \frac{\partial A}{\partial p_k}\frac{\partial B}{\partial q_k}\right)$ for functions $A(\mathbf{q},\mathbf{p})$, $B(\mathbf{q},\mathbf{p})$. Thus the algebra of observables ${\mathcal O}_{F}$ under the Poisson bracket $\left\{\cdot,\cdot \right\}_{(P,S)}$ reproduces precisely the algebra of functions $F(\mathbf{q},\mathbf{p})$ under the phase space Poisson bracket $\left\{ \cdot,\cdot \right\}_{(q,p)}$. 

\subsection{Ensembles on phase space}
\label{EPS}

We now consider the extension of the approach of ensembles on configuration space to a phase space description. To do this, we first need to introduce a probability density on phase space, which we denote by $\varrho(\mathbf{q},\mathbf{p})$ to distinguish it from the probability $P(\mathbf{q})$ on configuration space introduced previously. Moreover, we introduce the corresponding canonically conjugate variable, which we denote by $\sigma(\mathbf{q},\mathbf{p})$. We require that $\varrho \geq 0$ and $\int d\omega\,\varrho=1$, where $d\omega=d\mathbf{q}d\mathbf{p}$ is the phase-space measure. 

\subsubsection{Equations of motion}
\label{subsubsec:EqMotion}

We assume once more that the equations of motion are derived from an ensemble Hamiltonian $\mathcal H[\varrho,\sigma]$ according to
\begin{eqnarray}
	\frac{\partial \varrho}{\partial t} &=& \left\{ \varrho,{\mathcal H} \right\}_{(\varrho,\sigma)}  = \frac{\delta{\mathcal H}}{\delta \sigma}\\
	\frac{\partial \sigma}{\partial t} &=& \left\{ \sigma,{\mathcal H} \right\}_{(\varrho,\sigma)}  =-\frac{\delta{\mathcal H}}{\delta \varrho},
\end{eqnarray}
where now the Poisson bracket is defined by $\{ \mu,\nu \}_{(\varrho,\sigma)}=\int d\omega\; \left(\frac{\delta \mu}{\delta \varrho}\frac{\delta \nu}{\delta \sigma} - \frac{\delta \mu}{\delta \sigma}\frac{\delta \nu}{\delta \varrho}\right)$ for functionals $\mu[\varrho,\sigma]$, $\nu[\varrho,\sigma]$.

Before we introduce the explicit form of the ensemble Hamiltonian, let us look for appropriate equations of motion for $\varrho$ and $\sigma$. Consider again a single non-relativistic particle under the action of a  potential with the standard phase-space Hamiltonian
$
	H =\frac{\mathbf{p}^{2}}{2M}+V \label{PSH}
$
and the corresponding Lagrangian $\mathscr{L}=\frac{\mathbf{p}^{2}}{2M}-V.$ Conservation of probability requires that the Liouville equation
\begin{equation}
	\frac{\partial\mathcal{\varrho}}{\partial t}+\left\{ \varrho,H\right\}_{(q,p)} =0\label{liouville}
\end{equation}
is satisfied.
Furthermore, in analogy to the approach of ensembles on configuration space, we will identify $\sigma$ with the action in phase space\footnote{We point out that the extension of ensembles on configuration space to ensembles on phase space introduces some redundancy in the description of the classical system because the momentum $\mathbf{p}$ can also be expressed in terms of the action. As shown below, we can address this issue with an appropriate choice of $\sigma$.}. This assumption provides us with a second equation,
\begin{equation}
	\frac{d\sigma}{dt}=\frac{\partial\sigma}{\partial t}+\left\{ \sigma,H\right\}_{(q,p)} =\mathscr{L}.\label{Sl}
\end{equation}
The Hamiltonian and the Lagrangian given above lead to the following two linear first-order partial differential equations for $\varrho$ and $\sigma$, 
\begin{align}
	\frac{\partial\mathcal{\varrho}}{\partial t}+\nabla_{q}\varrho\cdot\frac{\mathbf{p}}{M}-\nabla_{p}\varrho\cdot\nabla_{q}V & =0,\label{ro2}\\
	\frac{\partial\sigma}{\partial t}+\nabla_{q}\sigma\cdot\frac{\mathbf{p}}{M}-\nabla_{p}\sigma\cdot\nabla_{q}V & =\frac{\mathbf{p}^{2}}{2M}-V.\label{s2}
\end{align}
These equations can be derived from the ensemble Hamiltonian
\begin{align}
	&\mathcal{H}_{C}[\rho,\sigma] =\nonumber\\
 &\int d\omega\,\mathcal{\varrho}\left[\left(V-\frac{\mathbf{p}^{2}}{2M}\right)+\nabla_{q}\sigma\cdot\frac{\mathbf{p}}{M}-\nabla_{p}\sigma\cdot\nabla_{q}V\right],\label{Hphasespace}
\end{align}
as one can check by direct computation. 

Eqs. (\ref{ro2}) and (\ref{s2}) are uncoupled equations that can be solved separately for each of the canonically conjugate variables $\varrho$ and $\sigma$. While the Liouville equation (\ref{ro2}) is perhaps more fundamental in that it describes the evolution of the density in phase space, which is usually the quantity of interest, we will pay close attention to the choice of $\sigma$ and introduce a procedure that selects a particular type of solution for Eq. (\ref{s2}); i.e., that fixes $\sigma$. This choice will play a crucial role in the formulation of the theory, in particular when we address in the next section the issue of observables and their proper interpretation. 

As is well known
\cite{LandauLifshitz1976}, the classical action can be represented by 
\begin{equation}
	\sigma=\int \, \left(\mathbf{p}\cdot d\mathbf{q}-Hdt\right), \label{sigmafunctional}
\end{equation}
that is, as a \textit{functional} expressed as an integral over the classical trajectory of a particle. However, this way of expressing $\sigma$ is not appropriate for a theory of ensembles on phase space: the classical action $\sigma$ should be expressed as a \textit{function} rather than a functional.  
We can find a different way of expressing $\sigma$ by taking into consideration that the first order partial differential equation for $\sigma$, Eq. (\ref{s2}), is equivalent to the system of ordinary differential equations \cite{Schuh68,CourantHilbert1962vol2} 
\begin{eqnarray}
dt&=&\frac{dq_i}{(p_i/M)}=-\frac{dp_i}{(\partial V/\partial q_i)},\qquad i=1,2,3\nonumber\\
&=&\frac{d\sigma}{(|\mathbf{p}|^2/2M - V)}.\label{s2ODEs}
\end{eqnarray}
To see this, notice that Eq. (\ref{s2ODEs}) leads to the following equalities,
\begin{equation}
d\mathbf{q}=\frac{\mathbf{p}}{M}\,dt,\quad d\mathbf{p}=-\nabla_q V \, dt,\quad d\sigma=\left(\frac{|\mathbf{p}|^2}{2M}-V\right)dt,\label{s2systemODEs}
\end{equation}
but we also have
\begin{eqnarray}
    d\sigma &=& \frac{\partial \sigma}{\partial t}dt+\nabla_{q}\sigma\cdot d\mathbf{q}+\nabla_{p}\sigma\cdot d\mathbf{p}\nonumber\\
    &=& \left[\frac{\partial\sigma}{\partial t}+\nabla_{q}\sigma\cdot\frac{\mathbf{p}}{M}-\nabla_{p}\sigma\cdot\nabla_{q}V\right]dt.\label{dsigmaH2}
\end{eqnarray}
If we now set the right-hand side of the third equality of Eqs. (\ref{s2systemODEs}) equal to the right hand side of Eq.(\ref{dsigmaH2}), we get Eq. (\ref{s2}), as required.

Using the third and then the first equalities of Eqs. (\ref{s2systemODEs}), we can express $d\sigma$ as
\begin{eqnarray}
d\sigma &=& \left(\frac{|\mathbf{p}|^2}{2M}-V\right)dt
=\left[\frac{|\mathbf{p}|^2}{M}-\left(\frac{|\mathbf{p}|^2}{2M}+V\right)\right]dt \nonumber\\
&=&  \mathbf{p}\cdot d\mathbf{q} -\left(\frac{|\mathbf{p}|^2}{2M}+V\right)dt,\label{s2dsigma}
\end{eqnarray}
which we recognize as an equation for the differential of the action, cf. \cite{LandauLifshitz1976}. From Eq. (\ref{s2dsigma}) it follows that
\begin{equation}
\nabla_q \sigma = \mathbf{p},\qquad \nabla_p \sigma = 0,\label{nablasofsigma}
\end{equation}
as well as the equation that is satisfied by $\sigma$,
\begin{equation}
\frac{\partial \mathbf{\sigma}}{\partial t} + \frac{|\nabla \sigma|^2}{2M}+V=0, \label{HJEsigma}
\end{equation}
which is the Hamilton-Jacobi equation, as one would expect. 

One can think of the procedure that we have followed as equivalent to determining the solutions of Eq. (\ref{s2}) by projecting the classical system to configuration space. Nevertheless, we have introduced this particular approach because it generalizes to the case of hybrid systems.

There is an alternative procedure that does not require this projection and also leads to a $\sigma$ that has the required properties. To derive it, express $\varrho$ in the form
\begin{equation}
	\varrho(\mathbf{q},\mathbf{p},t)=\int d\omega_0 \, w(\mathbf{q}_0,\mathbf{p}_0) \, \varrho^{[tr]}(\mathbf{q},\mathbf{p},t;\mathbf{q}_0,\mathbf{p}_0),\label{rhogeneral}
\end{equation}
where $d\omega_0=dq_0 dp_0$, $w(\mathbf{q}_0,\mathbf{p}_0) \ge 0$, $\int d\omega_0 \, w(\mathbf{q}_0,\mathbf{p}_0)=1$, and the density $\varrho^{[tr]}$ corresponding to a single trajectory is given by
\begin{eqnarray}
	\varrho^{[tr]}(\mathbf{q},\mathbf{p},t;\mathbf{q}_0,\mathbf{p}_0)&=&\delta(\mathbf{q}-\mathbf{Q}(\mathbf{q}_0,\mathbf{p}_0,t))\\
	&~&\times\; \delta(\mathbf{p}-\mathbf{P}(\mathbf{q}_0,\mathbf{p}_0,t))\nonumber\label{rho1trajectory}
\end{eqnarray}
where $\mathbf{Q}$ and $\mathbf{P}$ are trajectories that only depend on the initial conditions $\mathbf{q}_0$ and $\mathbf{p}_0$ and the time $t$, satisfying
\begin{equation}
	\mathbf{Q}(\mathbf{q}_0,\mathbf{p}_0,t=0))=\mathbf{q}_0,\quad \mathbf{P}(\mathbf{q}_0,\mathbf{p}_0,t=0))=\mathbf{p}_0.
\end{equation}
Now choose $\sigma$ to be of the form
\begin{eqnarray}
	\sigma(\mathbf{q},\mathbf{p},t;\mathbf{q}_0,\mathbf{p}_0)&=&\eta(\mathbf{q},\mathbf{p})\\
	&~&+H(\mathbf{q},\mathbf{p})\left[\tau(\mathbf{q},\mathbf{p})-\tau(\mathbf{q}_0,\mathbf{p}_0)-t\right],\nonumber\label{sigmaqpt}
\end{eqnarray}
where $\eta(\mathbf{q},\mathbf{p})$ is a function that needs to be determined from Eq. (\ref{s2}) and $\tau$ satisfies $\{\tau,H\} = 1$. Then, as shown in Appendix \ref{Appendix_sigmaqpt}, the equalities
\begin{equation}
	\varrho \, \left(\nabla_q \sigma - \mathbf{p}\right)=0,\qquad \varrho \, \nabla_p \sigma = 0,\label{nablasofsigma2}
\end{equation}
are always valid for the $\sigma$ defined in Eq. (\ref{sigmaqpt}).

The main results of this section are Eqs. (\ref{ro2})-(\ref{s2}) for $\varrho$ and $\sigma$ and the derivation of general forms for $\sigma$ that always satisfy Eqs. (\ref{nablasofsigma2}).

\subsubsection{Observables, generators and ensemble averages}
\label{ObsGenEnsAveEPS}

While the functional form of the ensemble Hamiltonian of Eq. (\ref{Hphasespace}) might seem surprising, it turns out that it can be derived directly from van Hove's Hilbert space representation of the generators of contact transformations\footnote{We will look more closely at the van Hove representation when we discuss the Hilbert space formulation of classical particles in section \ref{vHHS}} \cite {VanHove1951}, as we show in Appendix \ref{sec:obsandvanHove}. Guided by this observation, we introduce the following \textit{general procedure} to construct the set of observables: Given a function $F(\mathbf{q},\mathbf{p})$ in phase space, we define the corresponding observable in the approach of ensembles on phase space by
\begin{eqnarray}
	{\mathcal O}_{F}[\varrho,\sigma] &=& \int d\omega\,\mathcal{\varrho}\left[\left(F-\mathbf{p} \cdot \nabla_p F \right) - \left\{F,\sigma \right\} \right]\nonumber\\
	&=&	    
	\int d\omega\,\mathcal{\varrho}\left[\left(F-\mathbf{p} \cdot \nabla_p F \right) \right.\nonumber\\
 & &\quad- \left.\left(\nabla_q F \cdot  \nabla_p \sigma - \nabla_p F \cdot  \nabla_q \sigma\right) \right]. \label{EPSObs}
\end{eqnarray}

As in the previous case, we are dealing with a Hamiltonian formalism and we have an algebra of observables that is defined via the Poisson brackets $\{\cdot, \cdot\}_{(\varrho,\sigma)}$. Thus, the ${\mathcal O}_{F}[\varrho,\sigma]$ of Eq. (\ref{EPSObs}) will play both the role of observables and the role of a generators. 

We now make use of either of the special solutions for $\sigma$ that we derived in the previous section and in particular apply the results of Eq. (\ref{nablasofsigma2}) to the calculation of the  numerical value of ${\mathcal O}_{F}$. This leads to	
\begin{eqnarray}
    {\mathcal O}_{F}'[\varrho,\sigma] 
    &:=& \left.{\mathcal O}_{F}[\varrho,\sigma] \right|_{\nabla_q \sigma = \mathbf{p}, \nabla_p \sigma = 0} \nonumber\\
    &=& \int d\omega\,\mathcal{\varrho}\left[\left(F-\mathbf{p} \cdot \nabla_p F \right)\right. \nonumber\\
    & &- \left.\left(\nabla_q F \cdot  \nabla_p \sigma - \nabla_p F \cdot  \nabla_q \sigma\right) \right|_{\nabla_q \sigma = \mathbf{p},\nabla_p \sigma = 0} \nonumber\\
    &=& \int d\omega\,\mathcal{\varrho} \, F,\label{OnsEPSAve}
\end{eqnarray}
where the notation ${\mathcal O}_{F}'$ is introduced to indicate that the observable is calculated using Eqs. (\ref{nablasofsigma2}). Thus, with the procedures for fixing $\sigma$ that we introduced in the previous section, \textit{the numerical value of the observable is always equal to the average of its corresponding phase space function $F(\mathbf{q},\mathbf{p})$}. 

\subsubsection{Observables and conservation of probability}

Just as in the case of observables defined for the approach of ensembles on configuration space, we require that the changes in $\varrho$ induced by the observables defined in Eq. (\ref{EPSObs}) preserve both the normalization and positivity of the probability. This is indeed the case, and the proofs are similar to the ones given in section \ref{ECSobsProb} for the case of ensembles on configuration space.

\subsubsection{Lie algebra of observables}

As a consequence of the definition of classical observables of Eq. (\ref{EPSObs}), the Poisson bracket for classical observables ${\mathcal O}_{F}[\varrho,\sigma]$ is \textit{isomorphic} to the phase space Poisson bracket,
\begin{equation}
	\left\{ {\mathcal O}_{A},{\mathcal O}_{B}\right\}_{(\varrho,\sigma)}={\mathcal O}_{\left\{ A,B \right\}_{(q,p)} }.\label{EPSiso1}
\end{equation}
The calculation is given in Appendix \ref{sec:Appisoepsps}.

\subsection{Hilbert space formulations of classical mechanics using van Hove operators}
\label{vHHS}

Koopman \cite{Koopman31} and von Neumann \cite{Neumann1930} were the first to show that one may formulate classical mechanics in Hilbert space. Much later, van Hove derived a unitary representation of the group of contact transformation \cite{VanHove1951} which provides the basis of the Hilbert space formulation of classical mechanics that we present here. For previous applications of van Hove operators in this context, see references \cite{Bondar_Gay-Balmaz_Tronci_PRA2019,Gay_Balmaz_2020,Gay-Balmaz2022a} .

\subsubsection{Equations of motion}
\label{EoMvH}

In the Hilbert space formulation of classical mechanics, the states are given by phase-space valued wavefunctions $\phi(\mathbf{q},\mathbf{p},t)$, with the inner product defined by 
\begin{equation}
\left\langle \phi\right|\left.\chi\right\rangle =\int d\omega\,\phi^{*}\chi.
\end{equation}

There is some freedom in the choice of the equation of motion for $\phi$, with different but related equations leading to the same classical dynamics \cite{Klein2017}. In van Hove's approach, the dynamics is given by the Schr\"{o}dinger-like equation
\begin{equation}
	i\hbar \frac{\partial \phi}{\partial t}=\hat{{\mathcal O}}_H\phi,\label{vHSE}
\end{equation}
where the Hamiltonian operator for a non-relativistic particle acting on $\phi$  is most conveniently written as
\begin{eqnarray}
	\hat{{\mathcal O}}_H\phi &=& \left[\left(\frac{\mathbf{p}^{2}}{2M}-V(\mathbf{x})\right)\right.\nonumber\\
	 &~& \quad \left. + i\hbar\left( \nabla_q V \cdot \nabla_p - \frac{\mathbf{p}}{M} \cdot \nabla_q  \right)\right]\phi. \label{covariant liouvillian}
\end{eqnarray}

We want to point out that the appearance of $\hbar$ in Eqs. (\ref{vHSE}) and (\ref{covariant liouvillian}) is not linked to any quantization procedure: the set of unitary representations of the group of contact transformations of van Hove \cite{VanHove1951} constitutes a continuous family of representations labeled by a real parameter that he calls $\alpha$ and we have set this $\alpha=1/\hbar$ to simplify the equations.

If we introduce Madelung variables $\phi=\sqrt{\varrho}e^{i\sigma/\hbar}$ for the classical wave function and calculate $\bar{\phi}\, i\hbar \frac{\partial \phi}{\partial t}=\bar{\phi}\,\hat{{\mathcal O}}_H\phi$, the real and imaginary parts of Eq. (\ref{vHSE}) read
\begin{eqnarray}
	\frac{\partial\mathcal{\varrho}}{\partial t}&=&\nabla_{q}V\cdot\nabla_{p}\varrho  -\frac{\mathbf{p}}{M}\cdot\nabla_{q}\varrho,\label{rovH}\\
	\varrho \, \frac{\partial\sigma}{\partial t} & =& \varrho \left[ \frac{\mathbf{p}^{2}}{2M}-V+\nabla_{q}V\cdot\nabla_{p}\sigma-\frac{\mathbf{p}}{M}\cdot\nabla_{q}\sigma \right],\label{svH}
\end{eqnarray}
which coincide with the equations for the dynamics of ensembles on phase space given by Eqs. (\ref{ro2}) and (\ref{s2}). 

In particular, this means that \textit{the solutions $\varrho$, $\sigma$ that are valid for classical systems described by ensembles on phase space are also solutions for the van Hove formulation by setting $\phi=\sqrt{\varrho}\,e^{i\sigma/\hbar}$} \footnote{with some pathological exceptions where the square root of $\varrho$ is not defined, like Klimontovich distributions $\varrho=\delta(\mathbf{q}-\mathbf{q}(t))\delta(\mathbf{p}-\mathbf{p}(t)) $ where $\mathbf{q}(t) $ and $\mathbf{p}(t)$ obey Hamiltonian equations of motion.}. As Eqs. (\ref{ro2}) and (\ref{s2}) are equivalent to Eqs. (\ref{rovH}) and (\ref{svH}), the approach that we developed for ensembles on phase space gives us a way of fixing the phase of the classical wave function; see the results on $\sigma$ given in section \ref{subsubsec:EqMotion}. 

\subsubsection{Observables, generators and ensemble averages}

In the Hilbert space approach using van Hove operators, a phase space function $F(\mathbf{q},\mathbf{p})$ is represented by an operator $\hat{{\mathcal O}}_F$, and the action of this operator on a classical wavefunction is given by
\begin{eqnarray}
\hat{{\mathcal O}}_F \phi &=& \left[\left( F-\mathbf{p} \cdot \nabla_p F \right)+ i\hbar\left( \nabla_q F \cdot \nabla_p - \nabla_p F \cdot \nabla_q  \right) \right]\phi \nonumber\\
&=& \left( F-\mathbf{p} \cdot \nabla_p F \right) \phi + i\hbar  \{F,\phi \}_{\{q,p\}}. \label{vanHoveOp} 
\end{eqnarray}
One can check that the Hamiltonian operator, Eq. (\ref{covariant liouvillian}), results from 
evaluating Eq. (\ref{vanHoveOp}) with the classical particle Hamiltonian $H=\frac{\mathbf{p}^{2}}{2M}+V$.

It is straightforward to calculate the relation between the expectation values $\langle \phi | \hat{{\mathcal O}}_F | \phi \rangle$ of the van Hove operators and the observables ${\mathcal O}_F[\varrho,\sigma]$ for ensembles on phase space. Using again the polar decomposition of the classical wave function, $\phi=\sqrt{\varrho}\,e^{i\sigma/\hbar}$, we find that
\begin{equation}
    \langle \phi | \hat{{\mathcal O}}_F | \phi \rangle = {\mathcal O}_F[\varrho,\sigma] ,
\end{equation}
where ${\mathcal O}_F[\varrho,\sigma]$ is given by Eq. (\ref{EPSObs}). When $\sigma$ satisfies Eqs. (\ref{nablasofsigma}) and (\ref{HJEsigma}), then
\begin{equation}
    \langle \phi | \hat{{\mathcal O}}_F | \phi \rangle = \langle \phi | \hat{{\mathcal O}}'_F | \phi \rangle = \int d\omega \, \varrho F(\mathbf{q},\mathbf{p})
\end{equation}
and the numerical value of $\langle \phi | \hat{{\mathcal O}}_F | \phi \rangle$ is precisely the ensemble average of $F(\mathbf{q},\mathbf{p})$ and it is independent of $\hbar$, as expected, where the operator $\hat{{\mathcal O}}'_F$ is defined by.
\begin{equation}
    \hat{{\mathcal O}}'_F  =  F(\mathbf{q},\mathbf{p}).
\end{equation}

\subsubsection{Observables and conservation of probability}

As in the previous two approaches, we must determine that the changes in $\varrho$ induced by the observables defined in Eq. (\ref{vanHoveOp}) preserve both the normalization and positivity of the probability. This is straightforward in the Hilbert space formulation. First we note that the probability density  $\varrho=|\phi|^2$ is non-negative from its very definition. Additionally, as the operators defined by Eq. (\ref{vanHoveOp}) are Hermitian in the Hilbert space of phase-space valued wavefunctions, it follows that the infinitesimal unitary transformations generated by $\hat{{\mathcal O}}_F$ will not change the normalization of $\varrho=|\phi|^2$. 

\subsubsection{Lie algebra of van Hove operators}

As was already pointed out by van Hove, the commutator for  operators $\hat{{\mathcal O}}_A$ is \textit{isomorphic} to the phase space Poisson bracket,
\begin{equation}
	\frac{1}{i\hbar}\left[\hat{{\mathcal O}}_A,\hat{{\mathcal O}}_B\right]={\mathcal O}_{\left\{ A,B \right\}_{(q,p)} }.\label{isovHps}
\end{equation}
A proof is given in Appendix \ref{APPisovHps}.

\subsubsection{Absence of an uncertainty principle}

We have seen that the solutions $\varrho$, $\sigma$ for ensembles in phase space are mapped to the classical wavefunction of van Hove, except for some pathological examples (e.g., where the square root of $\varrho$ is not well-defined). In particular, localized classical solutions for ensembles on configuration space that approximate delta functions are also solutions for van Hove's classical mechanics in Hilbert space. However, the commutator algebra of the operators of van Hove is isomorphic to the Poisson algebra of functions in phase space, so that we have 
\begin{eqnarray}
	&&\!\!\!\![\hat{{\mathcal O}}_q,\hat{{\mathcal O}}_p]\nonumber\\&&=\hat{{\mathcal O}}_q \hat{{\mathcal O}}_p - \hat{{\mathcal O}}_p \hat{{\mathcal O}}_q \nonumber\\
	&&= \left(q+i\hbar\frac{\partial}{\partial p}\right)\left(-i\hbar\frac{\partial}{\partial q}\right)-\left(-i\hbar\frac{\partial}{\partial q}\right)\left(q+i\hbar\frac{\partial}{\partial p}\right)\nonumber\\
	&&= i\hbar \,\label{vHxpc}
\end{eqnarray}
in the one-dimensional case.

This may seem puzzling since $[\hat{{\mathcal O}}_q,\hat{{\mathcal O}}_p]=i\hbar$ seems to be in contradiction with the existence of localized solutions. Nevertheless, as we show below, \textit{there is no contradiction here because no uncertainty relation can be obtained from Eq. (\ref{vHxpc})}.

To understand this, it will be useful to start by reviewing one of the standard ways of deriving the uncertainty relation \cite{LandauLifshitz1977}. Consider the ket
\begin{equation}
	|\alpha\rangle=(\hat{{\mathcal O}}_q + i\lambda \hat{{\mathcal O}}_p)|\phi\rangle ,
\end{equation}
where $\lambda$ is an arbitrary real parameter. We assume for simplicity that the mean values of the position and momentum for the state $|\phi\rangle$ are zero. Calculate
\begin{eqnarray}
\langle\alpha|\alpha\rangle &=& \langle\phi|(\hat{{\mathcal O}}_q - i\lambda \hat{{\mathcal O}}_p)(\hat{{\mathcal O}}_q + i\lambda \hat{{\mathcal O}}_p)|\phi\rangle\nonumber\\
	&=& \langle\phi|\hat{{\mathcal O}}_q \hat{{\mathcal O}}_q |\phi\rangle - \lambda \hbar  +\lambda^2\langle\phi|\hat{{\mathcal O}}_p \hat{{\mathcal O}}_p |\phi\rangle \ge 0. \label{alpha2}
\end{eqnarray}
As the last expression in Eq. (\ref{alpha2}) is quadratic in $\lambda$, the condition that it be non-negative leads to
\begin{equation}
\langle\phi|\hat{{\mathcal O}}_q \hat{{\mathcal O}}_q |\phi\rangle \langle\phi|\hat{{\mathcal O}}_p \hat{{\mathcal O}}_p |\phi\rangle \ge \left(\frac{\hbar}{2}\right)^2.
\end{equation}
However, such a derivation does not go through for van Hove operators. There is a \textit{crucial difference} between the operators $\hat{Q}$, $\hat{P}$ of quantum mechanics and the van Hove operators $\hat{{\mathcal O}}_q$, $\hat{{\mathcal O}}_p$ because
\begin{eqnarray}
	\hat{Q} \hat{Q} &=& \hat{Q^2}, \qquad \,\,\quad \hat{P} \hat{P} = \hat{P^2},\\
	\hat{{\mathcal O}}_q \hat{{\mathcal O}}_q &\ne& \hat{{\mathcal O}}_{q^2}, \qquad \hat{{\mathcal O}}_p\hat{{\mathcal O}}_p \ne \hat{{\mathcal O}}_{p^2}.	
\end{eqnarray}
As a matter of fact, we have
\begin{eqnarray}
	\hat{{\mathcal O}}_q \hat{{\mathcal O}}_q &=& \left(q+i\hbar \frac{\partial}{\partial p}\right)^2 = q^2+2i\hbar q \frac{\partial}{\partial p} - \hbar^2 \frac{\partial^2}{\partial p^2}, \\
	\hat{{\mathcal O}}_p \hat{{\mathcal O}}_p &=& \left(-i\hbar \frac{\partial}{\partial q}\right)^2 = -\hbar^2\frac{\partial^2}{\partial q^2},
\end{eqnarray}
which shows that $\hat{{\mathcal O}}_q \hat{{\mathcal O}}_q$ and $\hat{{\mathcal O}}_p \hat{{\mathcal O}}_p$ are not even van Hove operators as they involve second derivatives, and all van Hove operators have first derivatives only. As a consequence, they are not associated with \textit{any} classical observables. \textit{The derivation of the uncertainty relation requires interpreting $\langle\phi|\hat{{\mathcal O}}_q\hat{{\mathcal O}}_q |\phi\rangle$ and $\langle\phi|\hat{{\mathcal O}}_p\hat{{\mathcal O}}_p |\phi\rangle$ as expectation values of the square of the position and the square of the momentum, respectively, and this fails with the van Hove operators}.

As a result, there is no uncertainty relation for the van Hove formulation of classical mechanics, despite the isomorphism between commutators of van Hove operators and Poisson brackets of functions in phase space. To put it in more technical terms: The set of van Hove observables $\hat{{\mathcal O}}_F$ does not form a product algebra. That is, the product of two van Hove observables $\hat{{\mathcal O}}_F \hat{{\mathcal O}}_G$ is not necessarily a van Hove observable. Given two arbitrary van Hove observables $\hat{{\mathcal O}}_F$ and $\hat{{\mathcal O}}_G$, the only general way to get a third observable is through their commutator  $\frac{1}{i\hbar}[\hat{{\mathcal O}}_F,\hat{{\mathcal O}}_G]=\hat{{\mathcal O}}_{\{F,G\}}$. 

\subsection{Galilean invariance in the three approaches to classical mechanics}
\label{GalInv}

Since Galilean invariance plays an important role in non-relativistic systems, it is of interest to look at how this symmetry is implemented in the different approaches to classical mechanics that we have considered. We will later look at this in the context of hybrid systems, where complications can arise due to the interactions between classical and quantum systems.

A realization of the Galilei algebra is given by the generators of translations $\Pi_{i}$, rotations $L_{i}$, boosts $G_i$ and time translation $H$. These operators satisfy the Lie algebra
\begin{align}
\{H,\Pi_i\} &= 0, &  \{H,L_i\} &= 0,  \\
\{\Pi_i,\Pi_j\} &=0, &
\{L_i,\Pi_j\} &= \varepsilon_{ijk}\Pi_k\nonumber\\
\{L_i,L_j\} &= \varepsilon_{ijk}L_k, & \{L_i,G_j\} &= \varepsilon_{ijk}G_k,\nonumber\\
\{G_i,G_j\} &= 0, &
\{G_i,\Pi_j\} &= M\delta_{ij},\nonumber\\ \{G_i,H\} &= \Pi_i, \nonumber
\end{align}
where the choice of brackets $\{\cdot,\cdot\}$ will depend on the model of classical mechanics that is being considered; i.e., Poisson brackets $\{\cdot,\cdot\}_{q,p}$ when the generators are represented by functions of phase space, commutators $\frac{1}{i\hbar}[\cdot,\cdot]$ when the generators are represented by operators in Hilbert space, etc. As is well known \cite{Finkelstein1973}, the representation in terms of functions in phase space is 
\begin{eqnarray}
H=\frac{1}{2M}|\mathbf{p}|^2&,& \qquad \Pi_i=p_i, \\
\qquad L_i=\varepsilon_{ijk}x_j p_k&,&\qquad G_i=Mq_i-tp_i. \nonumber
\end{eqnarray}
We have presented rules to map phase space functions $F(\mathbf{q},\mathbf{p})$ to observables for all three approaches. These rules provide an algebra that is isomorphic to the one of functions in phase space. Therefore, it is sufficient to write down the generators of the Galilean group in terms of functions in phase space and then apply the corresponding rules to get the generators for ensembles on configuration space, ensembles on phase space and van Hove's Hilbert space formulation, respectively. 

We find that Galilean invariance can be explicitly realized in all the approaches that we consider. This procedure leads to the following generators, presented in the table:
\vfill
\onecolumngrid
\begin{center}
\begin{table}[h!]\caption{Explicit form of the generators for the three approaches.} \label{tab1}
\begin{center}
\renewcommand{\arraystretch}{1.5}
\begin{tabular}{|m{2em}||m{16em}|m{16em}|m{16em}|} 
\hline
		& Configuration space ensembles & Phase space ensembles & Hilbert Space (van Hove operators) \\ 
		& $P(\mathbf{q}),~S(\mathbf{q})$ & $\varrho(\mathbf{q},\mathbf{p}),~\sigma(\mathbf{q},\mathbf{p})$ & $\psi(\mathbf{q},\mathbf{p})$ \\ 
		& $\{\cdot,\cdot\}_{P,S}$ & $\{\cdot,\cdot\}_{\varrho,\sigma}$ & $[\cdot,\cdot]$ \\[3pt]				
		\hline
		\hline
		$\Pi_i$ & $\int d\mathbf{q} \, P \left(\frac{\partial S}{\partial q_i}\right)$ & $\int d\omega\,\mathcal{\varrho} \left(\frac{\partial \sigma}{\partial q_i}\right) $ & $-i\hbar \frac{\partial}{\partial q_i}$  \\[3pt] 
		\hline
		$L_i$ &  $\int d\mathbf{q} \, P \left(\varepsilon_{ijk}q_j\frac{\partial S}{\partial q_k}\right)$ & $\int d\omega\,\mathcal{\varrho}\varepsilon_{ijk}\left(q_{j}\frac{\partial \sigma}{\partial q_{k}}-\frac{\partial \sigma}{\partial p_{j}}p_k\right)$ & $-i\hbar\varepsilon_{ijk}\left(q_{j}\frac{\partial}{\partial q_{k}}-p_k\frac{\partial }{\partial p_{j}}\right)$  \\[3pt] 
		\hline
		$G_i$ & $\int d\mathbf{q} \, P  \left( M q_i - t \frac{\partial S}{\partial q_i} \right)$ & $\int d\omega\,\mathcal{\varrho}\left(Mq_{i}-M\frac{\partial \sigma}{\partial p_{i}}-t\frac{\partial \sigma}{\partial q_{i}}\right)$ & $Mq_{i}+i\hbar\left(M\frac{\partial}{\partial p_{i}}+t\frac{\partial}{\partial q_{i}}\right)$  \\[3pt] 
		\hline
		$H$ & $\int d\mathbf{q} \, P \left( \frac{1}{2M}|\nabla_q S|^2\right)$ & $\int d\omega\,\mathcal{\varrho}\left(\frac{1}{M}\nabla_{q}\sigma\cdot\mathbf{p}-\frac{1}{2M}|\mathbf{p}|^{2}\right)$ & $ - i\hbar  \frac{1}{M}\mathbf{p} \cdot \nabla_q -\frac{1}{2M}|\mathbf{p}|^{2}$  \\[3pt] 
		\hline
	\end{tabular}
\end{center}
\end{table}
\end{center}\vfill\newpage
\twocolumngrid
\subsection{Equivalence of the three approaches}

To discuss the equivalence of the three approaches of representing statistical theories of classical mechanics, we consider the equivalence of their algebras and solutions.

\subsubsection{Isomorphism of the Lie algebras}
As we have already discussed, for each of the three approaches we have rules to map functions $F(\mathbf{q},\mathbf{p})$ in phase space to observables (i.e., ${\mathcal O}_{F}[P,S]$, ${\mathcal O}_{F}[\varrho,\sigma]$ and ${\hat{{\mathcal O}}}_{F}$). The Lie algebra of observables with respect to the corresponding brackets (i.e., $\{\cdot,\cdot\}_{(P,S)}$, $\{\cdot,\cdot\}_{(\varrho,\sigma)}$ and $[\cdot,\cdot]$) is in each case isomorphic to the Lie algebra for functions in phase space in terms of the Poisson brackets $\{\cdot,\cdot\}_{q,p}$. Hence,  all approaches are equivalent from an algebraic point of view. 

\subsubsection{Mapping of solutions}
As already pointed out at the end of section \ref{EoMvH}, the solutions $\varrho$, $\sigma$ for classical systems described by ensembles on phase space are also valid for the Hilbert space formulation by setting $\phi=\sqrt{\varrho}\,e^{i\sigma/\hbar}$. We restrict in both cases to solutions of $\sigma$ that satisfy Eq.(\ref{nablasofsigma2}). Thus, these two approaches are equivalent as far as the solutions are concerned.  

Equivalence to the approach of ensembles on configuration space can also be established by mapping the phase space density $\varrho$ to a mixture of probability densities $P$ according to \cite{HallReginatto2016}
\begin{equation}	
\varrho(\mathbf{q},\mathbf{p},t) = \int d \boldsymbol{\alpha} \; w(\boldsymbol{\alpha})P(\mathbf{q},t|\boldsymbol{\alpha}) \; \delta(\mathbf{p} - \nabla_q S(\mathbf{q},t;\boldsymbol{\alpha})).
\end{equation}
where $w(\boldsymbol{\alpha})$ is the probability of finding the mixture in the state labeled by $\boldsymbol{\alpha}$, $P(q,t|\boldsymbol{\alpha})$ is a conditional probability given $\boldsymbol{\alpha}$, and the set of parameters $\boldsymbol{\alpha}$ labeling both the elements of the mixture and a complete solution of the Hamilton-Jacobi equation \cite{LandauLifshitz1976} (a derivation following Ref. \cite{HallReginatto2016} is reproduced in Appendix \ref{fromEPStoECS}). In this way, the phase space density $\varrho(q,p)$ is mapped to a mixture of configuration space ensembles and both approaches are equivalent. 

\section{Three models for interacting hybrid classical-quantum systems}

To describe a mixed quantum-classical system, we add a quantum particle to the classical particle and allow them to interact. We first describe how this is done for ensembles on configuration space and then adapt this procedure to the other two approaches. Here and in what follows, we use ($\mathbf{q}$,$\mathbf{p}$) to describe the phase-space coordinates of the classical particle and $\mathbf{x}$ to denote the position of the quantum particle.

\subsection{Hybrid model in the approach of ensembles on configuration space}
\label{HECS}

We have given an account of the description of a classical particle using ensembles on configuration space in section \ref{ECS}. Before we introduce hybrid classical-quantum ensembles, we present a summary of the ensemble description of a quantum particle \cite{HallReginatto2016}. 

The ensemble Hamiltonian for a quantum particle of mass $m$ is given by 
\begin{equation}
	{\mathcal H}_{Q}[P,S] = \int d\mathbf{x}\; P \left( \frac{|\nabla_x S|^2}{2m} + \frac{\hbar^2}{4P^2}\frac{|\nabla_x P|^2}{2m} + V(\mathbf{x})\right). \label{CQEH_ECS}
\end{equation}
This ensemble Hamiltonian leads to the equations
\begin{eqnarray}
	\frac{\partial S}{\partial t} &=& -\frac{|\nabla_x S|^2}{2m}  + \frac{\hbar^{2}}{2m}\frac{\nabla_x^{2}\sqrt{P}}{\sqrt{P}} -  V, \label{HJE_ECS_QM}\\
	\frac{\partial P}{\partial t} &=& -\nabla \cdot \left( P\frac{\nabla S}{m} \right). \label{CQE_ECS}
\end{eqnarray}
If we introduce a wavefunction and write it in terms of Madelung variables, \mbox{$\psi=\sqrt{P}e^{iS/\hbar}$}, Eqs. (\ref{HJE_ECS_QM})-(\ref{CQE_ECS}) are equivalent to the complex Schr\"{o}dinger equation
\begin{equation}	i\hbar\frac{\partial\psi}{\partial t}=-\frac{\hbar^{2}}{2m}\nabla^{2}\psi+V\psi.
\end{equation}

Given an operator $\hat{F}$ in the Schr\"{o}dinger representation, the quantum observable in the approach of ensembles on configuration space is defined by the expectation value of the corresponding
operator,
\begin{equation}
	{\mathcal O}^{QM}_{\hat{F}}=\int d\mathbf{x}d\mathbf{x}'\,\sqrt{P(\mathbf{x})P(\mathbf{x}')}e^{i[S(\mathbf{x})-S(\mathbf{x}')]/\hbar}\left\langle \mathbf{x}'\right|\hat{F}\left|\mathbf{x}\right\rangle .\label{quantumO}
\end{equation}
which differs from the definition (\ref{classicalO}) of the corresponding classical observables \cite{HallReginatto2016}.
The fundamental importance of the definition of quantum observable of Eq. (\ref{quantumO}) derives from the fact that the Poisson bracket for quantum observables is \textit{isomorphic} to the commutator in Hilbert space,
\begin{equation}
	\left\{ {\mathcal O}^{QM}_{\hat{M}},{\mathcal O}^{QM}_{\hat{N}}\right\} _{(P,S)}={\mathcal O}^{QM}_{[\hat{M},\hat{N}]/i\hbar}.\label{iso2}
\end{equation}

\subsubsection{Equations of motion and Galilean invariance for hybrid systems}

We now can construct a hybrid theory in configuration space by letting $P$ and $S$ depend on the joint space $(\mathbf{q},\mathbf{x})$ and postulating a Hamiltonian that is the sum of the free classical and quantum Hamiltonians plus an interaction term. We also require that the resulting hybrid model is Galilean invariant, which restricts the possible choices for the interaction terms to a scalar potential $V(|\mathbf{q}-\mathbf{x}|)$. It can be shown that adding a rotationally and translationally invariant interaction is sufficient to guarantee Galilean invariance of the theory \cite{HallReginatto2016}. Hence, the full hybrid Hamiltonian can be written as
\begin{eqnarray}
{\mathcal H}_{CQ}[P,S] &=&
\int d\mathbf{q}\,d\mathbf{x}\, P \Biggl[ \frac{|\nabla_q S|^2}{2M} + \frac{|\nabla S_x|^2}{2m} \\
&&\qquad\qquad+ \frac{\hbar^2}{4P^2}\frac{|\nabla_x P|^2}{2m} + V(|\mathbf{q}-\mathbf{x}|)\Biggr].\nonumber
\end{eqnarray}
The resulting equations of motion are 
\begin{align}
	& \frac{\partial S}{\partial t} + \frac{|\nabla_q S|^2}{2M} +\frac{|\nabla_x S|^2}{2m}  - \frac{\hbar^{2}}{2m}\frac{\nabla_x^{2}\sqrt{P}}{\sqrt{P}} + V=0, \label{HJE_HECS}\\
	& \frac{\partial P}{\partial t} +\nabla_q \cdot \left( P\frac{\nabla_q S}{M} \right) + \nabla_x \cdot \left( P\frac{\nabla_x S}{m} \right) = 0. \label{CE_HECS}
\end{align}

\subsubsection{Conservation of probability}

One can show that the ensemble Hamiltonian of Eq.(\ref{CQEH_ECS}) preserves the positivity and the normalization of the probability $P(\mathbf{q},\mathbf{x})$ \cite{HallReginatto2016}. The probabilities for the classical and quantum sectors are given by marginalization,
\begin{equation}
    P_C(\mathbf{q})=\int d \mathbf{x} \, P(\mathbf{q},\mathbf{x}), ~~~~~ P_Q(\mathbf{x})=\int d \mathbf{q} \, P(\mathbf{q},\mathbf{x}),
\end{equation}
and are therefore always positive. 

\subsubsection{Observables of classical and quantum subsystems}

The functionals of Eqs. (\ref{classicalO}) and (\ref{quantumO}) that represent classical and quantum observables, respectively, have the same functional form as before, except that now $P$ and $S$ can be functions of both $\mathbf{q}$ and $\mathbf{x}$. As in the purely classical case, observables equal average values.

The classical observable corresponding to a phase space function $F(\mathbf{q},\mathbf{p})$ is defined by 
\begin{equation}
{\mathcal O}_{F}=\int d\mathbf{q}d\mathbf{x}\,P(\mathbf{q},\mathbf{x})F(\mathbf{q},\nabla S(\mathbf{q},\mathbf{x})).\label{classicalOHyb}
\end{equation}and the quantum observable corresponding to an operator $\hat{F}$ is given by 
\begin{eqnarray}
    {\mathcal O}^{QM}_{\hat{F}}&=&\int d\mathbf{q}d\mathbf{x}d\mathbf{x}'\,\sqrt{P(\mathbf{q},\mathbf{x})P(\mathbf{q},\mathbf{x}')}\\
    & &\qquad\quad\times e^{i[S(\mathbf{q},\mathbf{x})-S(\mathbf{q},\mathbf{x}')]/\hbar}\left\langle \mathbf{x}'\right|\hat{F}\left|\mathbf{x}\right\rangle .\label{quantumOHyb} \nonumber
\end{eqnarray}

\subsection{Hybrid model in the approach of ensembles on phase space} 
\label{HEPS}

We can construct a hybrid model where the classical particle is described by  phase-space coordinates with a procedure similar to the one that we used in the last subsection.  That is, we add the ensemble Hamiltonians of the classical particle and the quantum particle together with an interaction term. However, there is a subtlety involved here: contrary to the previous case, one cannot simply require the interaction term to be rotationally and translationally invariant to guarantee Galilean invariance \cite{Bermudez_Manjarres_2021,Bermudez_Manjarres2023phase}.

\subsubsection{Equations of motion and Galilean invariance}

To derive appropriate equations of motion, we add a Galilean invariant interaction term to the sum of the ensemble Hamiltonians of the classical and quantum free particles. 

The Galilean invariant ensemble Hamiltonian for a hybrid system of two particles interacting via a potential takes the form \cite{Bermudez_Manjarres2023phase}
\begin{eqnarray}
    &&\mathcal{H}_{CQ}[\rho,\sigma] = \label{CQEH_EPS}\\
    &&\int d\omega\, d\mathbf{x} \, \mathcal{\varrho} \left[\nabla_{q}\sigma\cdot\frac{\mathbf{p}}{M}-\frac{\mathbf{p}^{2}}{2M} + \frac{|\nabla_x \sigma|^2}{2m} + \frac{\hbar^2}{4\varrho^2}\frac{|\nabla_x \varrho|^2}{2m}  \right.\nonumber\\
    && \qquad\qquad\qquad\qquad\left. + V(|\mathbf{q}-\mathbf{x}|) - \nabla_p\sigma \cdot \nabla_q V(|\mathbf{q}-\mathbf{x}|)\right].\nonumber
\end{eqnarray}
The equations of motion are
\begin{align}
    & \frac{\partial\mathcal{\varrho}}{\partial t}+\nabla_{q}\varrho\cdot\frac{\mathbf{p}}{M}-\nabla_{p}\varrho\cdot\nabla_{q}V +\nabla_x \cdot \left( \varrho\frac{\nabla_x \sigma}{m} \right) =0,\label{EM1_EPS2}\\
    & \frac{\partial\sigma}{\partial t}+\nabla_{q}\cdot\frac{\mathbf{p}}{M}-\nabla_{p}\sigma\cdot\nabla_{q}V -\frac{\mathbf{p}^{2}}{2M} \label{EM2_EPS2}\\
    &\qquad\qquad\qquad\quad+\frac{|\nabla_x \sigma|^2}{2m}  - \frac{\hbar^{2}}{2m}\frac{\nabla_x^{2}\sqrt{\varrho}}{\sqrt{\varrho}} +  V =0.\nonumber
\end{align}
 These equations are precisely the Madelung form \cite{Gay_Balmaz_2020} of the Hybrid system given in \cite{Bondar_Gay-Balmaz_Tronci_PRA2019}. However, while there is equivalence at the level of equations, it is not clear to what extent the interpretations of the formalisms match. One reason is the differences in the definitions of physical quantities (such as the densities associated with the classical and quantum sectors and the definition of observables/generators). Another reason is the different ways of handling the phase of the hybrid wavefunction. Approaches to solving Eqs. (\ref{EM1_EPS2}) and (\ref{EM2_EPS2}) are discussed in Appendix \ref{specialEPSsolutions}. 

\subsubsection{Conservation of probability}

One can show that the ensemble Hamiltonian of Eq. (\ref{CQEH_EPS}) preserves positivity and the normalization of the probability $\varrho(\mathbf{q},\mathbf{p},\mathbf{x})$. The probabilities for the classical and quantum sectors are given by marginalization,
    \begin{equation}		\varrho_C(\mathbf{q},\mathbf{p})=\int d \mathbf{x} \, \varrho(\mathbf{q},\mathbf{p},\mathbf{x}), ~~~~~ \varrho_Q(\mathbf{x})=\int d \omega \, \varrho(\mathbf{q},\mathbf{p},\mathbf{x}),
\end{equation}
and are therefore always positive. 

\subsubsection{Observables of classical and quantum subsystems}

The functionals that represent classical and quantum observables have the same functional form as before, except that now $\varrho$ and $\sigma$ can be functions of $\mathbf{q}$, $\mathbf{p}$ and $\mathbf{x}$.

The classical observable corresponding to a phase space function $F(\mathbf{q},\mathbf{p})$ is defined by 
\begin{eqnarray}
    {\mathcal O}_{F}[\varrho,\sigma] 
    &=& \int d\omega d\mathbf{x}\,\mathcal{\varrho}\left[\left(F-\mathbf{p} \cdot \nabla_p F \right)\right.\\ &&\qquad\qquad\left.
    - \left(\nabla_q F \cdot  \nabla_p \sigma - \nabla_p F \cdot  \nabla_q \sigma\right) \right].\nonumber
\end{eqnarray}
In Appendix \ref{specialEPSsolutions}, we show that we can limit ourselves to the set of solutions of the hybrid equations that satisfy Eq. (\ref{nablasofsigma2}) without restricting the space of solutions. Thus, we restrict to this class of solutions, in which the numerical value of the observables equals the average value of the corresponding phase space function, 
\begin{equation}
    {\mathcal O}_{F}'[\varrho,\sigma] 
    := \left.{\mathcal O}_{F}[\varrho,\sigma] \right|_{\nabla_q \sigma=\mathbf{p}, \nabla_p \sigma=0} =
    \int d\omega d\mathbf{x}\,\mathcal{\varrho} \, F.\label{classicalOHybEPS}
\end{equation}

The quantum observable corresponding to an operator $\hat{F}$ is given by 
\begin{eqnarray}
    {\mathcal O}^{QM}_{\hat{F}}&=&\int d\omega d\mathbf{x}d\mathbf{x}'\,\sqrt{P(\mathbf{q},\mathbf{p},\mathbf{x})P(\mathbf{q},\mathbf{p},\mathbf{x}')}\nonumber\\
    & &\quad\times e^{i[S(\mathbf{q},\mathbf{p},\mathbf{x})-S(\mathbf{q},\mathbf{p},\mathbf{x}')]/\hbar}\left\langle \mathbf{x}'\right|\hat{F}\left|\mathbf{x}\right\rangle .\label{quantumOHy_OP}
\end{eqnarray}
As in the purely classical case, observables equal average values (provided we follow the rules given above).

\subsection{Hybrid model in the Hilbert space approach}
\label{HvHHS}

For the Hilbert space approach, we follow a procedure similar to those for the previous models. We introduce a hybrid model by adding the Hamiltonian operators of the classical particle, the quantum particle, and an interaction term.

\subsubsection{Equations of motion and Galilean invariance}

The Galilean invariant Hamiltonian operator acting on a wavefunction $\psi(\mathbf{q},\mathbf{p},\mathbf{x})$ for a hybrid system of two particles interacting via a potential is given by
\begin{eqnarray}
    i\hbar\frac{\partial\psi}{\partial t}&=& \left[\frac{\mathbf{p}^{2}}{2M} + i\hbar\left( \nabla_q V \cdot \nabla_q - \frac{\mathbf{p}}{M} \cdot \nabla_p  \right)\right.\label{CQHO_HS}\\ 
   & &\qquad\qquad\qquad\quad \left.-\frac{\hbar^{2}}{2m}\nabla_x^{2}+V(|\mathbf{q}-\mathbf{x}|)\right]\psi.\nonumber
\end{eqnarray}
This hybrid wave equation was first considered in Ref. \cite{Bondar_Gay-Balmaz_Tronci_PRA2019}, see also Ref. \cite{Bermudez_Manjarres_2021}. When the wavefunction $\psi$ is written in terms of Madelung variables, Eq. (\ref{CQHO_HS}) becomes equivalent to the equations for hybrid ensembles on phase space, Eqs. (\ref{EM1_EPS2}-\ref{EM2_EPS2})

\subsubsection{Conservation of probability}

The probability density is always non-negative as it is given by $\rho=|\psi|^2$. Furthermore, the Hamiltonian operator of Eq. (\ref{CQHO_HS}) is Hermitian, so its action preserves the normalization of the probability. 

The probabilities for the classical and quantum sectors are given by marginalization,
\begin{equation}	                 \varrho_C(\mathbf{q},\mathbf{p})=\int d \mathbf{x} \, \varrho(\mathbf{q},\mathbf{p},\mathbf{x}),\quad\varrho_Q(\mathbf{x})=\int d \omega \, \varrho(\mathbf{q},\mathbf{p},\mathbf{x}),
\end{equation}
therefore they are always positive. 

\subsubsection{Observables of classical and quantum subsystems}

The operators that represent classical observables are the van Hove operators of Eq. (\ref{vanHoveOp}). The operators that represent quantum observables are the usual quantum operators in the Schr\"{o}dinger representation in configuration space with coordinate $\mathbf{x}$.

\subsection{Comparison of hybrid theories}
\label{comparison}

When the equations of motion, states, and observables of two hybrid theories can be mapped to each other, we consider the two theories to be equivalent. In this section, we examine the equivalence or non-equivalence of the three hybrid approaches that we discuss in this paper.

\subsubsection{Equivalence of hybrid theories in Hilbert space and of ensembles on phase space}

We first consider the equivalence of hybrid theories in Hilbert space and of ensembles on phase space. 

To see that the equations of motion are the same, we write the wavefunction $\psi$ in Eq. (\ref{CQHO_HS}) in terms of Madelung variables. One can check that Eq. (\ref{CQHO_HS}) becomes equivalent to Eqs. (\ref{EM1_EPS2}) and (\ref{EM2_EPS2}). As the equations of motion are the same, the states in both hybrid theories will be the same, with the map from states of ensembles on phase space to wavefunctions given by  $\varrho, \sigma \rightarrow \psi=\sqrt{\varrho}e^{i \sigma/\hbar}$. Finally, for any operator $\hat{F}$ of the hybrid theory in Hilbert space, there is a corresponding observable ${\mathcal O}^{hybrid}_{\hat{F}}[\varrho,\sigma]=\langle \psi | \hat{F} | \psi \rangle_{\psi=\sqrt{\varrho}e^{i \sigma/\hbar}}$ for ensembles on phase space. 

\subsubsection{Non-equivalence of hybrids theories of ensembles on configuration space and on phase space} 
\label{classAndQuant}

To show the non-equivalence of the hybrids theories of ensembles on configuration space and on phase space, it is sufficient to provide a counter-example. We consider the energy and show that the corresponding observables for the two approaches are not equal to each other. It will be sufficient to focus on the terms proportional to $\left({|\nabla_x P|^2}/{P^2}\right)$ and $\left({|\nabla_x \varrho|^2}/{\varrho^2}\right)$ that appear as contributions from the quantum particle to the ensemble Hamiltonians and which lead to a Bohm quantum potential term in the equations of motion; see Eqs. (\ref{CQEH_ECS}) and (\ref{CQEH_EPS}).

Notice that the terms that are added to the ensemble Hamiltonians in the case of hybrid systems; i.e., 
\begin{equation}
	{\mathcal Q}^{ECS}[P]  = \frac{\hbar^2}{8m} \int d\mathbf{x} d\mathbf{q} \,P \left[  \frac{|\nabla_q P|^2}{P^2}\right]
\end{equation}
for the configuration space approach and
\begin{equation}
	{\mathcal Q}^{EPS}[\varrho] = \frac{\hbar^2}{8m} \int d\omega d\mathbf{x} \, \varrho \left[\frac{|\nabla_q \varrho|^2}{\varrho^2}\right]
\end{equation}
for the phase space approach, will also appear in the corresponding \textit{energy observables} and thus will contribute to the total energy of the hybrid system. Thus, non-equivalence between ${\mathcal Q}^{ECS}[P]$ and ${\mathcal Q}^{EPS}[\varrho]$ will lead to observable differences.

It is straightforward to show that ${\mathcal Q}^{ECS}[P]$ and ${\mathcal Q}^{EPS}[\varrho]$ are not equivalent in the general case. We write
\begin{equation}
	\varrho(\mathbf{q,p,x}) =: P(\mathbf{q,x})P(\mathbf{p|q,x}) 
\end{equation}
where in the last equality we used the product rule of probability theory to write $\varrho(\mathbf{q,p,x})$ as the product of a prior probability $P(\mathbf{q,x})$ and a conditional probability $P(\mathbf{p|q,x})$. Then we have 
\small
\begin{eqnarray}
	\nabla_x \varrho(\mathbf{x,p,q}) &=& \left[\nabla_x P(\mathbf{q,x})\right] P(\mathbf{p|q,x}) + P(\mathbf{q,x})\left[\nabla_x P(\mathbf{p|q,x})\right], \nonumber
 \end{eqnarray}
 and
 \begin{eqnarray}
	&& \!\!\!\!\!\!\!{\mathcal Q}^{EPS}[\varrho] \nonumber\\[0.3em]
 &=& \frac{\hbar^2}{8m} \int d\omega d\mathbf{x} \, \varrho \left[\frac{|\nabla_x \varrho|^2}{\varrho^2}\right]\\
	&=& \frac{\hbar^2}{8m} \int d\omega d\mathbf{x} \, P(\mathbf{q,x})P(\mathbf{p}|\mathbf{q,x}) \left[\frac{|\nabla_x P(\mathbf{q,x})|^2}{\left[P(\mathbf{q,x})\right]^2}\right.\nonumber\\
	&~& \left. \qquad\qquad\,\,\, +\frac{|\nabla_x P(\mathbf{\mathbf{p}|\mathbf{q,x}})|^2}{\left[P(\mathbf{p|q,x})\right]^2} + 2 \, \frac{\nabla_x P(\mathbf{q,x}) \cdot \nabla_x P(\mathbf{p|\mathbf{q,x}})}{P(\mathbf{q,x}) P(\mathbf{p|q,x})} \right]\nonumber\\
	&=& {\mathcal Q}^{ECS}[P] + \frac{\hbar^2}{8m} \int d\omega d\mathbf{x} \, P(\mathbf{q,x})P(\mathbf{p}|\mathbf{q,x}) \nonumber\\
	&~&  \qquad\qquad\times \left[ \frac{|\nabla_x P(\mathbf{\mathbf{p}|\mathbf{q,x}})|^2}{\left[P(\mathbf{p|q,x})\right]^2} + 2 \, \frac{\nabla_x P(\mathbf{q,x}) \cdot \nabla_x P(\mathbf{p|\mathbf{q,x}})}{P(\mathbf{q,x}) P(\mathbf{p|q,x})} \right]\nonumber\\
	&\neq& {\mathcal Q}^{ECS}[P]\nonumber ,
\end{eqnarray}
\normalsize
where we used the definition of ${\mathcal Q}^{ECS}[P] $ and $\int d\mathbf{p}\,P(\mathbf{p|q,x})=1$ in the third equality.

We conclude that the two hybrid theories, the one in phase space and the one in configuration space, are \textit{not} equivalent, with measurable differences of physical quantities like the energy.

\subsubsection{A condition that leads to a state in which the hybrid phase space and configuration space formulations are equivalent}

There is a case of interest in which the hybrid theories coincide. If no entanglement between the classical and quantum particles is assumed \footnote{These are states that satisfy $P({\mathbf{q,x}})= P_C({\mathbf{q}})P_Q({\mathbf{x}})$ and $S({\mathbf{q,x}})=S_C({\mathbf{q}})+S_Q({\mathbf{x}})$ in the case of ensembles in configuration space, with corresponding equations in the case of ensembles on phase space.}, we can write $\varrho$ as the product
	\begin{eqnarray}
		\varrho(\mathbf{q,p,x}) &=& \varrho_C({\mathbf{q,p}})P_Q(\mathbf{x})\\
  &=& P_C(\mathbf{q})\delta(\mathbf{p}-\nabla_q S_C({\mathbf{q}}))P_Q(\mathbf{x}),\nonumber
	\end{eqnarray}
In this case, 
	\begin{eqnarray}
		{\mathcal Q}^{EPS}[\varrho] &=& \frac{\hbar^2}{8m} \int d\omega d\mathbf{x} \, \varrho \left[\frac{|\nabla_x \varrho|^2}{\varrho^2}\right]\\
		&=& \frac{\hbar^2}{8m} \int d\omega d\mathbf{x} \, P_C(\mathbf{q})\delta(\mathbf{p}-\nabla_q S_C({\mathbf{q}}))\nonumber\\&& \qquad\qquad\quad\qquad\times P_Q(\mathbf{x}) \left[\frac{|\nabla_x P_Q(\mathbf{x})|^2}{\left[P_Q(\mathbf{x})\right]^2} \right] \nonumber\\
		&=& {\mathcal Q}^{ECS}[P]\,.	\nonumber
	\end{eqnarray}	
Thus, there is equivalence between all three theories considered in this paper when there is no entanglement between classical and quantum subsystems. We point out that in this case, the classical and quantum subsystems behave as independent subsystems (as long as there is no interaction term in the ensemble Hamiltonian that leads to entanglement), and this property is preserved under time evolution.

\section{Discussion} \label{sec:discussion}

In the first part of this paper, we presented three statistical descriptions of classical particles: ensembles on configuration space, ensembles on phase space, and a Hilbert space formulation using van Hove operators. In all of them, there is a natural way of defining observables and a corresponding Lie algebra (a functional Poisson algebra or a commutator algebra) that is isomorphic to the usual Poisson algebra of functions of position and momentum in phase space. Finally, we showed that these three descriptions of classical particles are equivalent; i.e., provide different representations of the same underlying statistical theory.

While the approach of ensembles on configuration space is known \cite{HallReginatto2016}, the second approach, utilizing ensembles on phase space, was only proposed recently \cite{Bermudez_Manjarres2023phase} but without providing a detailed description like the one provided in this paper. In particular, we have shown that the passage from ensembles on configuration space to ensembles on phase space requires some care, especially as it concerns the handling of the canonically conjugate variable identified with the action (see section \ref{subsubsec:EqMotion}), but no fundamental difficulties arise. The third approach that we considered, a Hilbert space description based on van Hove operators, has particularly interesting features. Unlike the Koopman-von Neumann formulation of classical mechanics in Hilbert space and other more recent developments, in our approach an observable and its corresponding generator are represented by the \textit{same} operator, defined by the rule introduced by van Hove. This leads to a consistent theory provided we identify the phase of the classical wavefunction with the classical action, using the same procedure as for ensembles on phase space, and also fix the phase of the wavefunction. When we do that, we find that the phase of the wavefunction does not appear in the expectation values of the van Hove operators, which are always equal to the average value of the phase space function associated with that operator. This Hilbert space formulation of classical mechanics provides an alternative to the well-known Koopman-von Neumann formulation. 

The formulation of classical mechanics using van Hove operators clarifies certain issues concerning the representation of classical observables in Hilbert space. This is of relevance to the question of ``classicality'' (that is, the question of determining the defining properties of a classical system). As already pointed out by van Hove, the commutator algebra of the operators is isomorphic to the usual Poisson algebra of functions in phase space. This implies that the operators associated with the classical position and momentum do \textit{not} commute, as is apparent from the calculation of Eq. (\ref{vHxpc}). Nevertheless, it is not possible to derive an uncertainty principle because the operators do not form a product algebra: while the commutator of two van Hove operators is another van Hove operator, the product of two van Hove operators is not necessarily a van Hove operator. Thus the widespread belief that operators in Hilbert space representing classical observables \textit{must} commute and that this property can be used to distinguish what is classical from what is quantum appears to be incorrect.

In the second part of the paper, the approaches are modified and extended to describe a hybrid system where a classical particle interacts with a quantum particle. For this step, it is necessary to establish appropriate forms of the classical-quantum interaction term and here the requirement of Galilean invariance provides crucial guidance \cite{Bermudez_Manjarres_2021, Bermudez_Manjarres2023phase}. The approach of ensembles on phase space and the Hilbert space approach lead to equivalent hybrid models, but they are not equivalent to the hybrid model of the approach of ensembles on configuration space. Thus we end up identifying two \textit{inequivalent} types of hybrid systems.  

The results that we have obtained for hybrid systems are of interest with regard to various ``no-go'' theorems about quantum systems interacting via a classical mediator. The motivation for them comes mostly from efforts to determine whether it is logically necessary to quantize gravity. It is known that such ``no-go'' theorems are in general model dependent \cite{HallReginatto2018}. The hybrid systems that we present in our paper provide concrete examples of inequivalent models that can be used to compute simple examples with the aim of testing the assumptions of the ``no-go'' theorems and their applicability.

The approach of linear operators acting on a classical or a hybrid wavefunction has a natural Hilbert space formulation. The other two approaches, ensembles on configuration space and on phase space, are Hamiltonian hydrodynamical formulations. If these mathematical structures are preserved, there should be no difficulty in considering generalizations beyond the scope of this paper.

For classical systems, the three approaches are physically equivalent. Whether one of them is preferred depends on the particular application; e.g. practical computations, approximation schemes. For hybrid systems, however, it is known that the different approaches proposed in the literature typically lead to different theories, usually leading to different predictions. This also holds true for the approach of ensembles on configuration space being not equivalent to the other two approaches. Thus, in the hybrid case it is ‘not just a matter of taste’ which approach to use. This is an aspect that still needs to be investigated. For example, it would be interesting to see which of the hybrid approaches provides a better approximation to a purely quantum system, as is often done in quantum chemistry, and, ultimately, which model holds ‘true’ when one of the sectors of the hybrid system is classical in nature. 

A detailed comparison to other hybrid models in the literature is beyond the scope of this paper and will be the topic of another publication.


\acknowledgements{One of the authors, M. Reginatto, is grateful for an invitation to the \textit{Quantum-classical interface in closed and open systems} meeting at the University of Surrey, where initial results of this paper were presented. The authors want to thank Cesare Tronci and Giovanni Manfredi for valuable discussions. S. Ulbricht acknowledges the funding by the Deutsche Forschungsgemeinschaft (DFG, German Research Foundation) under Germany’s Excellence Strategy—EXC 2123 QuantumFrontiers—390837967.}

\appendix
\section*{Appendices}

\section{An appropriate choice of $\sigma$}
\label{Appendix_sigmaqpt}

We consider, for the case in which the phase space of the system is two dimensional, a choice of $\sigma$ satisfying $\varrho ({\partial \sigma}/{\partial q}-p)=0$ and $\varrho{\partial \sigma}/{\partial p}=0$. It is straightforward to generalize to phase spaces of more dimensions.

As discussed in section \ref{subsubsec:EqMotion}, an arbitrary density $\varrho$ can be written in the form given by Eq. (\ref{rhogeneral}), which depends on the densities $\varrho^{[tr]}$ that describe single trajectories, given by Eq. (\ref{rho1trajectory}). We evaluate the Liouville equation, Eq. (\ref{ro2}), with the density of Eq. (\ref{rho1trajectory}). This leads to
\begin{equation}
	\left.\dot{Q}=\frac{p}{M}\right|_{p=P}\;,\qquad \left.\dot{P}=-\frac{\partial V}{\partial q}\right|_{q=Q},\label{eqstrajectories}
\end{equation}
where $Q$ and $P$ are the trajectories, which only depend on the initial conditions $q_0$ and $p_0$ and the time $t$, and satisfy  $Q(q_0,p_0,t=0)=q_0$ and $P(q_0,p_0,t=0)=p_0$.

Evaluate now the equation for $\sigma$, Eq. (\ref{s2}), multiplied with $\varrho$, with $\sigma$ as in Eq. (\ref{sigmaqpt}). Choosing the $\eta$ of Eq. (\ref{sigmaqpt}) so that it satisfies 
$ \left\{\eta,H\right\} - {p^2}/{2M} + V = 0$, we get the result
\begin{eqnarray}
	&~&{\varrho} \,\left[\frac{\partial {\sigma}}{\partial t} + \left\{{\sigma},H\right\} - \frac{p^2}{2M} + V \right]\\
	&~& \quad = {\varrho} \,\left[-H + \left\{\eta,H\right\} + H\left\{\tau,H\right\}- \frac{p^2}{2M} + V \right]\nonumber\\
	&~& \quad = 0\nonumber\label{stilde2a} 	
\end{eqnarray}
But we also have

\begin{eqnarray}
	&~&{\varrho} \,\left[\frac{\partial {\sigma}}{\partial t} + \left\{{\sigma},H\right\} - \frac{p^2}{2M} + V \right]\\ 
	&~& \quad = {\varrho} \,\left[-H + \frac{\partial {\sigma}}{\partial q}\frac{p}{M} 
	- \frac{\partial {\sigma}}{\partial p}\frac{\partial V}{\partial q} - \frac{p^2}{2M} + V \right]\nonumber\\
	&~& \quad = {\varrho} \,\left[\frac{\partial {\sigma}}{\partial q}\frac{p}{M} 
	- \frac{\partial {\sigma}}{\partial p}\frac{\partial V}{\partial q} - \frac{p^2}{M} \right] = 0.\nonumber
	\label{stilde2b} 	
\end{eqnarray}
The delta functions in ${\varrho}$ enforce motion along the trajectories that contribute to ${\varrho}$. In particular, along each of those trajectories, at any time $t$, we have $\dot{q}=\frac{p}{M}$ and $\dot{p}=-\frac{\partial V}{\partial q}$. Thus we can replace the right hand side of Eq. (\ref{stilde2b}) by
\begin{eqnarray}
	&~&{\varrho} \,\left[\frac{\partial {\sigma}}{\partial q}\dot{q}
	+ \frac{\partial {\sigma}}{\partial p}\dot{p} - p\dot{q} \right]\\
	&~& \quad = {\varrho} \,\left[\left(\frac{\partial {\sigma}}{\partial q}-p\right)\dot{q}
	+ \frac{\partial {\sigma}}{\partial p}\dot{p} \right] = 0.\nonumber
	\label{stilde2c} 	
\end{eqnarray}
Since this is valid for any choice of ${\varrho}$, no matter which trajectories and values of $\dot{q}$ and $\dot{p}$ are involved, it follows that
\begin{equation}
	{\varrho} \,\left(\frac{\partial {\sigma}}{\partial q}-p\right)=0,\qquad
	{\varrho} \,\frac{\partial {\sigma}}{\partial p}=0.\label{AppAderivsigmas}
\end{equation} 

As an example, consider the case of free fall, where $H=\frac{p^2}{2m}+mgq$. We choose 
\begin{equation}
	\eta=qp + \frac{p^3}{6m^2g},\qquad \tau=-\frac{p}{mg},
\end{equation}
which leads to
\begin{equation}
	{\sigma}=qp + \frac{p^3}{6m^2g}+\left(\frac{p^2}{2m}+mgq\right)\left(-\frac{p}{mg}+\frac{p_0}{mg}-t\right).\label{ts}
\end{equation}
At $t=0$ we have $p=p_0$, which means that the numerical value of the last term in parenthesis in Eq. (\ref{ts}) is zero. But $\tau$ evolves in time at the same rate as $t$ because $\{\tau,H\}=1$. This means that the numerical value of the last term in Eq. (\ref{ts}) is zero for \textit{all} times $t$. When we evaluate the derivatives of ${\sigma}$ over any trajectory, we obtain
\begin{equation}
	\frac{\partial {\sigma}}{\partial q} =  p, \qquad \frac{\partial {\sigma}}{\partial p} = 0.
\end{equation}
This implies the equalities of Eq. (\ref{AppAderivsigmas}).

\section{Observables for ensembles in phase space and their relation to van Hove's operators}
\label{sec:obsandvanHove}

The van Hove operator $\hat{{\mathcal O}}_F$ associated with the phase space function $F(\mathbf{q},\mathbf{p})$ acts on the classical wave function $\phi$ according to Eq. (\ref{vanHoveOp}). To define corresponding observables ${\mathcal O}_F$ for ensembles on phase space, we introduce Madelung variables, $\phi=\sqrt{\varrho}\,e^{i\sigma/\hbar}\,$ and evaluate the expectation value of the van Hove operator,
\begin{eqnarray}
	{\mathcal O}_F &:=& \langle \hat{{\mathcal O}}_F \rangle\nonumber\\ 
	&=& \int d\omega \varrho \left[ F- \nabla_p F\cdot\left(\mathbf{p}-\nabla_q\sigma  \right) - \nabla_q F\cdot\nabla_p\sigma  \right]\nonumber\\
	&~& \quad -\frac{ i \hbar}{2} \int d\omega \left(\nabla_p F\cdot\nabla_q \varrho  + \nabla_q F\cdot \nabla_p \varrho   \right)\nonumber\\
	&=& \int d\omega \varrho \left[ F- \nabla_p F\cdot \mathbf{p} - \left\{F, \sigma \right\}_{(q,p)} \right]
\end{eqnarray}
where in the last equality we integrated by parts assuming $\varrho \rightarrow 0$ at the boundaries. In particular, the choice $F(\mathbf{q},\mathbf{p})=
\frac{|\mathbf{p}|^2}{2m}+V$ leads to Eq. (\ref{Hphasespace}).

\section{The algebras of observables of ensembles on phase space and of functions on phase space}
\label{sec:Appisoepsps}

We show that the Lie algebras of observables of ensembles on phase space and of functions on phase space are isomorphic. For simplicity, we prove this result for the one dimensional case. We  have
\begin{eqnarray}
	{\mathcal O}_{F}[\varrho,\sigma] 
	&=& \int d\omega\,\mathcal{\varrho}\left[\left(F-p \frac{\partial F} {\partial p} \right) - \left\{F, \sigma \right\}_{(q,p)}\right], \nonumber\\
	\frac{\partial {\mathcal O}_F} {\partial \varrho} &=& \left(F-p \frac{\partial F} {\partial p} \right) - \left\{F, \sigma \right\}_{(q,p)}, \nonumber\\
	\frac{\partial {\mathcal O}_F} {\partial \sigma} &=& \frac{\partial } {\partial p}\left(\varrho  \frac{\partial F} {\partial q} \right) - \frac{\partial } {\partial q}\left(\varrho  \frac{\partial F} {\partial p} \right). 
\end{eqnarray}
We want to calculate the functional Poisson brackets between two observables, $\left\{ {\mathcal O}_F,{\mathcal O}_G \right\}_{(\rho,\sigma)}$, and show that it equals the observable ${\mathcal O}_{\{F,G\}}$ that corresponds to the function $\{F,G\}_{(q,p)}$, where the Poisson bracket of $F$ and $G$ is evaluated in phase space. This establishes the isomorphism between the two algebras. 

We need to calculate,
\small
\begin{eqnarray}
	&~& \left\{ {\mathcal O}_F,{\mathcal O}_G \right\}_{(\rho,\sigma)}\\
	&~& \quad = \int d\omega\; \left(\frac{\delta {\mathcal O}_F}{\delta \varrho}\frac{\delta {\mathcal O}_G}{\delta \sigma} - \frac{\delta {\mathcal O}_F}{\delta \sigma}\frac{\delta {\mathcal O}_G}{\delta \varrho}\right)\nonumber\\
	&~& \quad = \int d\omega\; \left[
	\left(F-p \frac{\partial F} {\partial p}  - \left\{F, \sigma \right\}_{(q,p)} \right)\right.\nonumber\\
	&~& \qquad \quad \times 
	\left. \left( \frac{\partial } {\partial p}\left(\varrho  \frac{\partial G} {\partial q} \right) - \frac{\partial } {\partial q}\left(\varrho  \frac{\partial G} {\partial p} \right) \right)\right] - F \leftrightarrow G \nonumber\\
	&~& \quad = \int d\omega\; \varrho \left[
	\frac{\partial } {\partial q} \left(F-p \frac{\partial F} {\partial p}  - \left\{F, \sigma \right\}_{(q,p)} \right) \frac{\partial G} {\partial p}\right. \nonumber\\ 
	&~& \qquad \quad \left. - \frac{\partial } {\partial p} \left(F-p \frac{\partial F} {\partial p}  - \left\{F, \sigma \right\}_{(q,p)} \right) \frac{\partial G} {\partial q}\right] - F \leftrightarrow G \nonumber
\end{eqnarray}	
\normalsize
where ``$F \leftrightarrow G$'' indicates that the previous expression is repeated interchanging $F$ and $G$, and in the last equality we integrated by parts assuming $\varrho \rightarrow 0$ at the boundaries. After quite a bit of algebra, one can show that
\small
\begin{eqnarray}
	\left\{ {\mathcal O}_F,{\mathcal O}_G \right\}_{(\rho,\sigma)}
	&=& \int d\omega\; \varrho \left[\{F,G\}_{(q,p)}\right.\\
	&~& \quad \left.-\frac{\partial } {\partial p}\{F,G\}_{(q,p)}-\{\{F,G\},S\}_{(q,p)}\right],\nonumber
\end{eqnarray}
\normalsize
as required. This shows that \textit{the Lie algebra of the ${\mathcal O}_F$ observables for ensembles on phase space is isomorphic to the Lie algebra of the functions $F(\mathbf{q},\mathbf{p})$ on phase space},
\begin{equation}
	\left\{ {\mathcal O}_F,{\mathcal O}_G \right\}_{(\rho,\sigma)}={\mathcal O}_{\{F,G\}_{(q,p)}}.
\end{equation}

\section{The algebras of van Hove operators and of observables of ensembles on phase space and functions on phase space}
\label{sec:isoepsps}

In this Appendix, we show that the commutator algebra of the van Hove operators is isomorphic to both the Poisson algebra of phase space functions and to the functional Poisson algebra of the observables defined for ensembles on phase space. 

\subsection{The commutator algebra of the van Hove operators is isomorphic to the Poisson algebra of phase space functions}
\label{APPisovHps}

In this subsection of the Appendix, we omit the subscript $(q,p)$ when we write the Poisson brackets $\{\cdot,\cdot\}_{(q,p)}$ of functions in phase space to simplify the notion. This cannot lead to confusion because only one type of Poisson brackets are needed here. 

The commutator of two van Hove operators acting on the classical wavefunction $\phi$ is given by
\begin{eqnarray}	
[\hat{{\mathcal O}}_F,\hat{{\mathcal O}}_G] \phi &=&  i\hbar\left[\{F,\Gamma_G\}-\{G,\Gamma_F\}\right]\phi \\
& &-\hbar^2 \left[\{F,\{G,\phi\}\}-\{G,\{F,\phi\}\}\right] ,\nonumber
\end{eqnarray}
where we introduced the notation $\Gamma_F=\left( F-\mathbf{p} \cdot \nabla_p F \right)$ and used the result
\begin{eqnarray}
    \hat{{\mathcal O}}_F \hat{{\mathcal O}}_G \phi &=& \Gamma_F \Gamma_G \phi + i\hbar\left[\Gamma_F  \{G,\phi\} + \Gamma_G \{F,\phi\}+\{F,\Gamma_G\} \phi\right] \nonumber\\
    &&-\hbar^2 \{F,\{G,\phi\}\}.
\end{eqnarray}
We use Jacobi's identity to write
\begin{eqnarray}
	\{F,\{G,\phi\}\}-\{G,\{F,\phi\}\}&=&\{F,\{G,\phi\}\}+\{G,\{\phi,F\}\}\nonumber\\
 &=&-\{\phi,\{F,G\}\}
\end{eqnarray}
which leads to 
\begin{equation}	[\hat{{\mathcal O}}_F,\hat{{\mathcal O}}_G] \phi =  i\hbar\left[\{F,\Gamma_G\}-\{G,\Gamma_F\}\right]\phi -\hbar^2 \{\{F,G\},\phi\} .
\end{equation}

The aim is to check that $[\hat{{\mathcal O}}_F,\hat{{\mathcal O}}_G] \phi$ equals
\begin{eqnarray}	&&i\hbar\hat{{\mathcal O}}_{\{F,G\}} \phi \\&&=  i\hbar\left[\{F,G\}-\mathbf{p} \cdot \nabla_p\{F,G\}\right]\phi -\hbar^2 \{\{F,G\},\phi\} .\nonumber 
\end{eqnarray}
As the terms proportional to $\hbar^2$ match, it is only necessary to examine the remaining terms. We have
\small
\begin{eqnarray}
	&&\!\!\!\!\!\!i\hbar\left[\{F,\Gamma_G\}-\{G,\Gamma_F\}\right]\nonumber\\&=&i\hbar\left[\left\{F,\sum_jp_j\frac{\partial G}{\partial p_j}\right\}-\left\{G,\sum_jp_j\frac{\partial F}{\partial p_j}\right\}\right]\nonumber\\
	&=& i\hbar\left[\sum_k\left(\frac{\partial F}{\partial q_k}\frac{\partial }{\partial p_k}\left(\sum_jp_j\frac{\partial G}{\partial p_j}\right)\right)\right.\nonumber\\ &&\qquad\qquad\qquad\qquad\qquad\qquad-\frac{\partial F}{\partial p_k}\frac{\partial }{\partial q_k}\left(\sum_jp_j\frac{\partial G}{\partial p_j}\right)\nonumber\\
	&&\qquad\qquad- \sum_k\left(\frac{\partial G}{\partial q_k}\frac{\partial }{\partial p_k}\left(\sum_jp_j\frac{\partial F}{\partial p_j}\right)\right)\nonumber\\&&\qquad\qquad\qquad\qquad\qquad\qquad+\left.\frac{\partial G}{\partial p_k}\frac{\partial }{\partial q_k}\left(\sum_jp_j\frac{\partial F}{\partial p_j}\right)\right]\nonumber\\
	&=& i\hbar\left[\sum_k\left(\frac{\partial F}{\partial q_k}\left(\frac{\partial G}{\partial p_k}+\sum_jp_j\frac{\partial^2 G}{\partial p_k \partial p_j}\right)\right)\right.\nonumber\\
 && \qquad\qquad\qquad\qquad\qquad\qquad-\frac{\partial F}{\partial p_k}\left(\sum_jp_j\frac{\partial^2 G}{\partial q_k \partial p_j}\right)\nonumber\\
	&&\qquad\qquad- \sum_k\left(\frac{\partial G}{\partial q_k}\left(\frac{\partial F}{\partial p_k}+\sum_jp_j\frac{\partial^2 F}{\partial p_j \partial p_k}\right)\right)\nonumber\\
&&\qquad\qquad\qquad\qquad\qquad\qquad+\left.\frac{\partial G}{\partial p_k} \left(\sum_jp_j\frac{\partial^2 F}{\partial q_k \partial p_j}\right)\right]\nonumber\\
	&=& i\hbar\left[\{F,G\}-\mathbf{p} \cdot \nabla_p\{F,G\}\right]
\end{eqnarray}
\normalsize

It follows then that
\begin{equation}
	[\hat{{\mathcal O}}_F,\hat{{\mathcal O}}_G] \phi = i\hbar\hat{{\mathcal O}}_{\{F,G\}} \phi,
\end{equation}
as required. 

This proves that \textit{the commutator algebra of the van Hove operators is isomorphic to the Poisson algebra of phase space functions}. Thus the formulation of classical mechanics in Hilbert space of van Hove is, in an algebraic sense, strictly equivalent to the usual one in phase space.

\subsection{The commutator algebra of the van Hove operators is isomorphic to the functional Poisson algebra of the observables defined for ensembles on phase space}
\label{isovHEPS}

In Appendix \ref{sec:obsandvanHove}, we defined ${\mathcal O}_F$, the observable of the theory of ensembles on phase space that corresponds to the phase space function $F(\mathbf{q},\mathbf{p})$, using the corresponding operator of van Hove's Hilbert space representation of classical mechanics. As the van Hove operators are Hermitian operators, we can write
\begin{equation}
    {\mathcal O}_F = \int d\omega \, \bar{\phi} \hat{{\mathcal O}}_F \phi = \int d\omega \, (\overline{\hat{{\mathcal O}}_F{\phi}})  \phi 
\end{equation}
We write the classical wave function using Madelung variables, $\phi=\sqrt{\varrho}\,e^{i\sigma/\hbar}\,$. Then the functional derivatives of ${\mathcal O}_F$ are given by
\begin{eqnarray}
\frac{\delta {\mathcal O}_F}{\delta \varrho}&=&\frac{\delta {\mathcal O}_F}{\delta \phi}\frac{\delta \phi}{\delta \varrho}+\frac{\delta {\mathcal O}_F}{\delta \bar{\phi}}\frac{\delta \bar{\phi}}{\delta \varrho}\\&=&\frac{\delta {\mathcal O}_F}{\delta \phi}\frac{\phi}{2 \varrho}+\frac{\delta {\mathcal O}_F}{\delta \bar{\phi}}\frac{ \bar{\phi}}{2 \varrho},\nonumber\\
	\frac{\delta {\mathcal O}_F}{\delta \sigma}&=&\frac{\delta {\mathcal O}_F}{\delta \phi}\frac{\delta \phi}{\delta \sigma}+\frac{\delta {\mathcal O}_F}{\delta \bar{\phi}}\frac{\delta \bar{\phi}}{\delta \sigma}\\&=&\frac{i}{\hbar}\left(\frac{\delta {\mathcal O}_F}{\delta \phi}\phi+\frac{\delta {\mathcal O}_F}{\delta \bar{\phi}} \bar{\phi}\right),\nonumber
\end{eqnarray}
and

\small
\begin{eqnarray}
	&& \!\!\!\!\!\!\!\!\left\{ {\mathcal O}_F,{\mathcal O}_G \right\}_{(\rho,\sigma)}\nonumber\\
	&=& \int d\omega\; \left(\frac{\delta {\mathcal O}_F}{\delta \varrho}\frac{\delta {\mathcal O}_G}{\delta \sigma} - \frac{\delta {\mathcal O}_F}{\delta \sigma}\frac{\delta {\mathcal O}_G}{\delta \varrho}\right)\nonumber\\
	&=& \frac{i}{\hbar}\int d\omega\; \left[\left(\frac{\delta {\mathcal O}_F}{\delta \phi}\frac{\phi}{2 \varrho}+\frac{\delta {\mathcal O}_F}{\delta \bar{\phi}}\frac{ \bar{\phi}}{2 \varrho}\right)\left(\frac{\delta {\mathcal O}_G}{\delta \phi}\phi+\frac{\delta {\mathcal O}_G}{\delta \bar{\phi}} \bar{\phi}\right)\right.\nonumber\\	
	&~& \qquad \qquad - \left.\left(\frac{\delta {\mathcal O}_F}{\delta \phi}\frac{\phi}{2 \varrho}+\frac{\delta {\mathcal O}_F}{\delta \bar{\phi}}\frac{ \bar{\phi}}{2 \varrho}\right)\left(\frac{\delta {\mathcal O}_G}{\delta \phi}\phi+\frac{\delta {\mathcal O}_G}{\delta \bar{\phi}} \bar{\phi}\right)\right]\nonumber\\	
	&=& \frac{1}{i \hbar}\int d\omega\; \left(\frac{\delta {\mathcal O}_F}{\delta \phi}\frac{\delta {\mathcal O}_G}{\delta \bar{\phi}}-\frac{\delta {\mathcal O}_F}{\delta \bar{\phi}}\frac{\delta {\mathcal O}_G}{\delta \phi}\right)\nonumber\\	
	&=& \frac{1}{i \hbar}\int d\omega\; \left((\overline{\hat{{\mathcal O}}_F{\phi}})  \hat{{\mathcal O}}_G\phi-\hat{{\mathcal O}}_F\phi(\overline{\hat{{\mathcal O}}_G{\phi}})\right)\nonumber\\	
	&=& \frac{1}{i \hbar}\int d\omega\; \bar{\phi}\left(\hat{{\mathcal O}}_F\hat{{\mathcal O}}_G-\hat{{\mathcal O}}_G\hat{{\mathcal O}}_F\right)\phi\nonumber\\	
	&=&  K_{[\hat{{\mathcal O}}_F,\hat{{\mathcal O}}_G]}					
\end{eqnarray}

Thus the functional Poisson Lie algebra of the ${\mathcal O}_F$ observables of the ensembles on phase space is isomorphic to the commutator Lie algebra of the van Hove operators $\hat{{\mathcal O}}_F$ in Hilbert space.

\section{Phase space densities and mixtures on configuration space ensembles}
\label{fromEPStoECS}

We summarize some results on how densities in phase space can be mapped to mixtures on configuration space   \cite{HallReginatto2016}. For simplicity, we consider a two-dimensional phase space. The generalization to more dimensions is straightforward. 

We write the phase space density (at some given time) in the form
\begin{equation}
	\varrho(q,p) = \int d q' d p' \; \varrho(q',p') \; \delta(q-q') \; \delta(p-p')
\end{equation}
and we carry out the change of coordinates
\begin{equation}\label{p0}
	p' = \frac{\partial S(q',\alpha)}{\partial q'}.
\end{equation}
where $S$ is a complete solution of the Hamilton-Jacobi equation with parameter $\alpha$. The only restriction on $S(q',\alpha)$ comes from the requirement that the coordinate transformation of Eq. (\ref{p0}) be invertible. We have
\begin{eqnarray}
	&&d q' d p' \varrho(q',p') \delta(p-p') \\
&&\qquad = d q' d \alpha \; \left| \frac{\partial^2 S}{\partial q' \partial \alpha} \right| \varrho(q',\partial S/\partial q') \delta(p - \partial S/\partial q'),\nonumber
\end{eqnarray}
which leads to
\begin{eqnarray}\label{varrho}
	&&\varrho(q,p)\\ &&= \int d q' d \alpha \; \left| \frac{\partial^2 S}{\partial q' \partial \alpha} \right| \; \varrho(q',\partial S/\partial q') \; \delta(q-q') \; \delta(p - \partial S/\partial q')\nonumber\\
	&&= \int d \alpha \; \left| \frac{\partial^2 S}{\partial q \partial \alpha} \right| \; \varrho(q,\partial S/\partial q) \; \delta(p - \partial S/\partial q). \nonumber
\end{eqnarray}

We now evaluate
\begin{eqnarray}
	\varrho(q) &=& \int d p \; \varrho(q,p) \nonumber\\
	&=&  \int d p \; d \alpha \; \left| \frac{\partial^2 S}{\partial q \partial \alpha} \right| \; \varrho(q,\partial S/\partial q) \; \delta(p - \partial S/\partial q) \nonumber\\
	&=&  \int d \alpha \; \left| \frac{\partial^2 S}{\partial q \partial \alpha} \right| \; \varrho(q,\partial S/\partial q) \nonumber\\
	& =: &  \int d \alpha \; w(\alpha) P(q|\alpha),
\end{eqnarray}
where the last line defines a pair of new probabilities, $w(\alpha)$ and $P(q|\alpha)$\footnote{We introduced the slash notation to indicate a conditional probability: $P(q|\alpha)$ is the probability for $q$ given a particular value of $\alpha$.}. Thus, we can set
\begin{equation}\label{wP}
	\left| \frac{\partial^2 S}{\partial q \partial \alpha} \right| \; \varrho(q,\partial S/\partial q) = w(\alpha)P(q|\alpha).
\end{equation}

It is possible to give explicit expressions for both $w(\alpha)$ and $P(q|\alpha)$. Integrating Eq. (\ref{wP}) with respect to $q$ on both sides leads to
\begin{equation}\label{appwalpha}
	w(\alpha) = \int dq \; \left| \frac{\partial^2 S}{\partial q \partial \alpha} \right| \; \varrho(q,\partial S/\partial q),
\end{equation}
where we used $\int d q \; P(q|\alpha)=1$ (see below). Using Eq. (\ref{wP}) again we get
\begin{eqnarray}\label{Pqalpha}
	P(q|\alpha) &=& \frac{1}{w(\alpha)}\left| \frac{\partial^2 S}{\partial q \partial \alpha} \right| \; \varrho(q,\partial S/\partial q)\\ &=& \frac{\left| \frac{\partial^2 S}{\partial q \partial \alpha} \right| \; \varrho(q,\partial S/\partial q)}{\int dq \; \left| \frac{\partial^2 S}{\partial q \partial \alpha} \right| \; \varrho(q,\partial S/\partial q)}.
\end{eqnarray}
Thus $w(\alpha)$ and $P(q|\alpha)$ are uniquely determined by $\varrho(q,p)$ and $S(q,\alpha)$.

Both $w(\alpha)$ and $P(q|\alpha)$ are non-negative and properly normalized, as we now show. Since the integrand of Eq. (\ref{appwalpha}) is non-negative, it follows that $w(\alpha) \geq 0$, as required. One can also show that $\int d\alpha \; w(\alpha)=1$, since
\begin{eqnarray}
	1 &=& \int dq dp \; \varrho(q,p) \\ &=&
 \int dq d\alpha \; \left| \frac{\partial^2 S}{\partial q \partial \alpha} \right| \; \varrho(q,\partial S/\partial q) = \int d\alpha \; w(\alpha), \nonumber
\end{eqnarray}
where in the second equality we carried out the transformation of Eq. (\ref{p0}) by replacing primed coordinates with unprimed coordinates. Finally, inspection of Eq. (\ref{Pqalpha}) shows that $P(q|\alpha) \geq 0$ and $\int d q \; P(q|\alpha)=1$.

Using Eq. (\ref{wP}), the expression for $\varrho(q,p)$, Eq. (\ref{varrho}), becomes
\begin{equation}
	\varrho(q,p) = \int d \alpha \; w(\alpha)P(q|\alpha) \; \delta(p - \partial S(q;\alpha)/\partial q).
\end{equation}
This shows that $\varrho(q,p)$ is indeed mapped to a mixture of configuration space ensembles.

Notice that the functions $P(q|\alpha)$ and $S(q,\alpha)$ have only been defined at a given instant of time. Given $P(q|\alpha)$ and $S(q,\alpha)$, one can derive the corresponding time-dependent expressions $P(q|\alpha,t)$ and $S(q,\alpha,t)$ by solving the equations of motion of the ensemble on configuration space, Eqs. (\ref{CE_HJE_ECS}), with $P(q|\alpha)$ and $S(q,\alpha)$ as initial conditions. This leads to an expression of the form
\begin{equation}
	\varrho(q,p,t) = \int d \alpha \; w(\alpha)P(q,t|\alpha) \; \delta(p - \partial S(q,t;\alpha)/\partial q).
\end{equation}

\section{Solving the equations for hybrid systems: Ensembles on phase space and Hilbert space formulation}
\label{specialEPSsolutions}

For simplicity, we consider the case of one dimensional classical and quantum particles. It is straightforward to generalize the argument to the case of more dimensions.

Consider Eqs. (\ref{EM1_EPS2})-(\ref{EM2_EPS2}), which describe a hybrid ensemble on configuration space. As these equations can be derived from the wave equation for a hybrid system in Hilbert space, Eq. (\ref{CQHO_HS}), via a Madelung transformation, it will be sufficient for the purposes of this Appendix to only sketch the derivation of solutions for the case of ensembles on phase space. The basic ideas (modulo technical issues) carry over to the derivation of solutions for the Hilbert space formulation. 

We associate with the density $\varrho(q,p,x,t)$ of the hybrid system a velocity field that determines (for at least a given time interval) its motion in time. In the case of a classical system, the velocity field can be identified with the particle trajectories, while in the case of a quantum system it will correspond to  Bohm trajectories \cite{Holland1993}. In the case of a hybrid system, it may be difficult to determine trajectories explicitly but this will not be needed here. 

For a single trajectory $\varrho^{[tr]}(q,p,x,t;q_0,p_0,x_0)$, we will have
\begin{eqnarray}
	\varrho^{[tr]}&=&\delta(q-Q(q_0,p_0,x_0,t))\delta(p-P(q_0,p_0,x_0,t))\nonumber\\
	&~& \quad \times \; \delta(x-X(q_0,p_0,x_0,t))\label{sigma1trajectoryh}
\end{eqnarray}
where $Q$, $P$ and $X$ are trajectories that only depend on the initial conditions $q_0$, $p_0$ and $x_0$ and the time $t$, satisfying
\begin{eqnarray}
Q(q_0,p_0,x_0,t=0) &=& q_0,\quad P(q_0,p_0,x_0,t=0) = p_0,\nonumber\\
X(x_0,p_0,x_0,t=0) &=& x_0.
\end{eqnarray}
If we evaluate Eq. (\ref{EM1_EPS2}) with the density of Eq. (\ref{sigma1trajectoryh}), it leads to
\begin{equation}
	\left.\dot{Q}=\frac{p}{M}\right|_{p=P}\;,\quad \left.\dot{P}=-\frac{\partial V}{\partial q}\right|_{q=Q}\;,\quad \left.\dot{X}=\frac{1}{m}\frac{\partial \sigma }{\partial x}\right|_{x=X}\qquad\label{eqstrajectoriesh}
\end{equation}
which are the equations of the trajectory.

We can now write any arbitrary density $\varrho(q,p,x,t)$ in the form
\begin{equation}
	{\varrho}=\int d\omega_0 dx_0 \, w(q_0,p_0,x_0) \; \varrho^{[tr]}(q,p,x,t;q_0,p_0,x_0)\label{rhogeneralh}
\end{equation}
where $d\omega_0=dq_0 dp_0$, $w(q_0,p_0,x_0) \ge 0$ and $\int d\omega_0 dx_0\, w(q_0,p_0,x_0)=1$. 

We now consider Eq. (\ref{EM2_EPS2}) with the density of Eq. (\ref{rhogeneralh}), and $\sigma$ given by
\begin{eqnarray}
	{\sigma}(q,p,x,t;q_0,p_0,x_0)&=&\eta(q,p,x,t)\\
	&~& +H(q,p)\left[\tau(q,p,x)-\tau(q_0,p_0,x_0)-t\right]\label{sigmahEPS}\nonumber
\end{eqnarray}
where $\eta(q,p,x,t)$ is a function that needs to be determined from Eq. (\ref{EM2_EPS2}) and $\tau$ satisfies 
$\{\tau,H\} = 1$.

If we evaluate Eq. (\ref{EM2_EPS2}) multiplied by $\varrho$ using the $\sigma$ of Eq. (\ref{sigmahEPS}), we get the following equation,
\begin{eqnarray}
	&&{\varrho} \,\left[\frac{\partial {\sigma}}{\partial t} + \left\{{\sigma},H\right\} - \frac{p^2}{2M} + V + \frac{1}{2M}\left(\frac{\partial {\sigma} }{\partial x}\right)^2 - \frac{\hbar^2}{2m}\frac{\partial^2 \sqrt{{\varrho}} }{\partial x^2} \right]\nonumber\\
	&~& = {\varrho} \,\left[ \frac{\partial \eta}{\partial t}+\frac{\partial {\sigma}}{\partial q}\frac{p}{M} 
	- \frac{\partial {\sigma}}{\partial p}\frac{\partial V}{\partial q} - \frac{p^2}{M} + \frac{1}{2M}\left(\frac{\partial {\sigma} }{\partial x}\right)^2\right.\nonumber\\
 &~& \quad \qquad - \left.\frac{\hbar^2}{2m}\frac{\partial^2 \sqrt{{\varrho}} }{\partial x^2} \right]\nonumber\\
	&~& = {\varrho} \,\left[ \frac{\partial \eta}{\partial t}+\left(\frac{\partial {\sigma}}{\partial q}-p\right)\dot{q} 
	+ \frac{\partial {\sigma}}{\partial p}\dot{p} + \frac{1}{2M}\left(\frac{\partial {\sigma} }{\partial x}\right)^2 - \frac{\hbar^2}{2m}\frac{\partial^2 \sqrt{{\varrho}} }{\partial x^2} \right]\nonumber\\
	&~& = 0 \label{stilde2ah}
\end{eqnarray}
where we used the fact that the delta functions in ${\varrho}$ enforce motion along the trajectories that contribute to the density ${\varrho}$, which implies $\dot{q}=\frac{p}{M}$ and $\dot{p}=-\frac{\partial V}{\partial q}$ at any given time $t$. Since the substitutions $\frac{p}{M}\rightarrow\dot{q}=$ and $-\frac{\partial V}{\partial q}\rightarrow\dot{p}$ are valid for any choice of ${\varrho}$, no matter which trajectories and values of $\dot{q}$ and $\dot{p}$ are involved, we must have
\begin{equation}\label{dsigmah}
	\varrho\left(\frac{\partial {\sigma}}{\partial q}-p\right)=0,\qquad \varrho \frac{\partial {\sigma}}{\partial p}=0. 
\end{equation} 

We now combine Eq. (\ref{stilde2ah}) and Eq. (\ref{dsigmah}) and derive the equation satisfied by $\eta$,
\begin{equation}
	{\varrho} \,\left[ \frac{\partial \eta}{\partial t} + \frac{1}{2M}\left(\frac{\partial (\eta+H\tau) }{\partial x}\right)^2 - \frac{\hbar^2}{2m}\frac{\partial^2 \sqrt{{\varrho}} }{\partial x^2} \right]=0
\end{equation}

\bibliography{
sn-article_v2.bib}

\end{document}